\documentclass[12pt]{article}

\usepackage{amsmath, amsthm, amscd, amsfonts, amssymb, graphicx, color}
\usepackage[bookmarksnumbered, colorlinks, plainpages]{hyperref}
\usepackage{pspicture,pst-all,multido}
\theoremstyle{definition}

\theoremstyle{remark}

\numberwithin{equation}{section}

\newcommand{\C}{\mathbb{C}}
\newcommand{\R}{\mathbb{R}}
\newcommand{\N}{\mathbb{N}}

\renewcommand{\d}{\operatorname{d}}
\newcommand{\e}{\operatorname{e}}
\renewcommand{\i}{\operatorname{i}}

\newcounter{envcount}%

{\vspace{\bigskipamount}\refstepcounter{envcount}\textbf{(\theenvcount)\enspace Asymptotic causality.}}%
  {\vspace{\bigskipamount}}

\newenvironment{Def}%
{\vspace{\bigskipamount}\refstepcounter{envcount}\textbf{(\theenvcount)\enspace Definition.}}%
  {\vspace{\bigskipamount}}

{\vspace{\bigskipamount}\refstepcounter{envcount}\textbf{(\theenvcount)\enspace Dependence on $c$.}}%
  {\vspace{\bigskipamount}}

{\vspace{\bigskipamount}\refstepcounter{envcount}\textbf{(\theenvcount)\enspace Example.}}%
  {\vspace{\bigskipamount}}

{\vspace{\bigskipamount}\refstepcounter{envcount}\textbf{(\theenvcount)\enspace Remarks.}}%
  {\vspace{\bigskipamount}}

\newenvironment{Rem}%
{\vspace{\bigskipamount}\refstepcounter{envcount}\textbf{(\theenvcount)\enspace Remark.}}%
  {\vspace{\bigskipamount}}

{\vspace{\bigskipamount}\refstepcounter{envcount}\textbf{(\theenvcount)\enspace  }}%
  {\vspace{\bigskipamount}}

\newenvironment{The}%
{\vspace{\bigskipamount}\refstepcounter{envcount}\textbf{(\theenvcount)\enspace Theorem.}\itshape}%
  {\vspace{\bigskipamount}\upshape}

{\vspace{\bigskipamount}\refstepcounter{envcount}\textbf{(\theenvcount)\enspace Theorem}\itshape}%
  {\vspace{\bigskipamount}\upshape}
  
\newenvironment{Pro}%
{\vspace{\bigskipamount}\refstepcounter{envcount}\textbf{(\theenvcount)\enspace Proposition.}\itshape}%
  {\vspace{\bigskipamount}\upshape}

\newenvironment{Cor}%
{\vspace{\bigskipamount}\refstepcounter{envcount}\textbf{(\theenvcount)\enspace Corollary.}\itshape}%
  {\vspace{\bigskipamount}\upshape}

\newenvironment{Lem}%
{\vspace{\bigskipamount}\refstepcounter{envcount}\textbf{(\theenvcount)\enspace Lemma.}\itshape}%
  {\vspace{\bigskipamount}\upshape}

\theoremstyle{definition}
\swapnumbers

\begin{document}
\setcounter{page}{1}
\pagenumbering{arabic}
\begin{center}
{\Large Achronal localization and representation of the causal logic from a conserved current {with an} application to the massive scalar boson}

\vspace{0.5cm}
Domenico P.L. Castrigiano$^{a1}$, Carmine De Rosa$^{b2}$,  Valter Moretti$^{b3}$(*)\\

\vspace{0.2cm}

$^a$Technische Universit\"at M\"unchen, Fakult\"at f\"ur Mathematik, M\"unchen, Germany\\ 
$^b$Dipartimento di Matematica, Universit\`a di Trento and TIFPA-INFN, Trento, Italy\\
(*) = corresponding author
\smallskip

{\it E-mail addresses}: {\tt   $^1$castrig\,\textrm{@}\,ma.tum.de, \: $^2$carmine.derosa@unitn.it\: $^3$valter.moretti@unitn.it}\\

\end{center}

\abstract
{Only recently the concept of achronal localization has been developed as the adequate frame for the description of the localizability of a relativistic quantum mechanical system.
Here} covariant achronal localizations are gained out of covariant conserved currents computing their flux passing through achronal surfaces. This general method  is applied to the  probability density currents with causal kernel regarding the massive scalar boson.  {As  (covariant) achronal localizations correspond  one-to-one to (covariant) representations of the causal logic, thus,} apparently for the first time, a covariant representation of the causal logic for an   elementary relativistic quantum mechanical system has been achieved. Similarly  a covariant family of representations of the causal logic is derived {from} the stress-energy tensor of the massive scalar boson.

{The construction of an achronal localization from a conserved current relies  on a version of the  divergence theorem for open sets with almost Lipschitz boundary. This result  is stated and proved in this work.}

\section{Introduction} 
$\qquad$
\textbf{Achronal localization.}  
 {Localizability of a relativistic quantum mechanical system concerns not only flat spacelike regions, but all  {\em achronal} regions of spacetime. One gains this basic insight  when studying thoroughly   the fact that localizability has to comply with causality. For a mathematical description a {\bf localization operator}  $T(\Delta)$ is assigned to every achronal Borel set $\Delta\subset\R^4$ {(see sec.\,\ref{NN}, Achronal sets)}. $T(\Delta)$ is a  a nonnegative  operator  bounded by $I$ acting on the  Hilbert space $\mathcal{H}$ of states. {The expectation value $\langle \phi, T(\Delta) \phi\rangle$
 is} regarded as} the probability of localization  in $\Delta$ of the system in the state represented by the unit vector $\phi \in \mathcal{H}$. 
 As argued further in \cite{C24}, the probabilities of localization regarding achronally separated regions {countably} add up and yield $1$ for every maximal achronal region. In summary $T$ defines  a normalized {\em positive operator valued measure} ({POVM}) on every maximal achronal set. {The map  $T$ which associates every achronal Borel set $\Delta\subset\R^4$ to $ T(\Delta)$ is called an {\bf achronal localization} (AL) of the system.} Moreover, by relativistic symmetry, $T$ is assumend to be Poincar\'e covariant {with respect to the unitary representation  of the Poincaré group corresponding to kinematical  transformations of the states}. {This type of AL $T$  is said to be a {\bf covariant} AL.}
\\
\hspace*{6mm} 
{\bf Heuristic.} One may imagine a \textbf{massive free particle} to be represented by a timelike straight world line. The particle is considered to be localized in a spacetime region   if its world line crosses the region. This simple heuristic picture of localizability  implies rather stringent achronal localization  \cite[sec.\,3.3]{C24}. Since the 
 maximal achronal sets are  those  sets, which are met by every timelike straight line just once,  one heuristically infers that a massive particle is localized in every maximal achronal region.
 Hence normalization of $T$ follows on this type of sets. The present results on the localization of the massive  scalar boson evidence the usefulness of this heuristical reasoning.
 \\
\hspace*{6mm} 
{The question is how to realize an  apparatus, which ascertains the presence of a particle in an achronal spacetime region $\Delta$. From a physical point of view it is reasonable to assume the region to be spacelike piecewise flat. Presume that localization in $3D$ Euclidean space is feasible. This implies by  relativistic symmetry that for every flat piece $\Gamma$ there is an apparatus suited to ascertain the presence of the particle in that region $\Gamma$. Being spacelike separated all these apparatus constitute one apparatus able to ascertain the presence of the particle in $\Delta$.}
\\
\hspace*{6mm} 
 {\bf Causality.} Achronal localization is causal: it satisfies  the so-called Einstein causality in the most direct interpretation, which  requires  that the probability of localization in {{\itshape any region of influence}} of the actual localization cannot be less than that in the region of actual localization. {Formal definitions and details} about  this causality condition (CC) and achronal localization are furnished in sec.\,\ref{SALCCC}.
\\
\hspace*{6mm} 
\textbf{Localization on spacelike Cauchy surfaces.} Just in order to meet the requirements of causal locali{zation, De Rosa and Moretti  \cite{DM24}} extended localization from flat spacelike regions, as up to then commonly considered, to proper spacetime {regions. They  study} thoroughly POVM on spacelike smooth Cauchy surfaces, which are a special case of maximal achronal sets.
The POVM are coherent in the sense that the localization operators assigned to a region contained in the intersection of two different spacelike smooth Cauchy surfaces coincide. An important result is that the  localization considered in \cite{DM24} satisfies a rather general causality requirement. 
{The results in \cite{DM24} were achieved by a suitable use of the divergence theorem for volume forms. It was done by  taking advantage of relatively recent advanced  results in Lorentzian geometry  concerning the extension of acausal manifolds with boundary to  spacelike smooth Cauchy surfaces.
These achievements actually straightforwardly   extend to {spacelike}  $C^1$ Cauchy surfaces.}

{In this work, we do the final step,
the construction of an achronal localization. It} goes along the lines of the construction of the Cauchy localization in \cite{DM24}. However there are  some technical difficulties, {which arise due to the facts (i)  that a maximal achronal set in general is not a Cauchy surface as it is not met by every lightlike  line  and (ii) that an achronal set in general is not $C^1$, which means that it is not the graph of a $C^1$ function. } {Achronal sets are  the graphs of Lipschitz functions, which in general are less regular than $C^1$ functions.}
\\
\hspace*{6mm} 
\textbf{Divergence theorem.} {The divergence theorem plays a crucial role in  the construction of concrete achronal localization notions.} {In order to overcome the difficulties} {pointed out above,}  we prove the divergence theorem for open bounded subsets of $\R^n$ with almost Lipschitz boundary such that the boundary has finite $(n-1)$-dimensional Hausdorff measure and the irregular points of the boundary are contained in a compact set of zero $(n-1)$-dimensional Minkowski content. This extension (\ref{MT}) of the divergence theorem  is new. In view of applications we add some remarks about how to verify the assumptions of the theorem. At the beginning of sec.\,4 it is expounded how (\ref{MT}) applies in showing the crucial technical result  (\ref{GMTA}).
\\
\hspace*{6mm}
\textbf{Flux through maximal achronal sets.} Using the prior  result the crucial result is proven that the future-directed flux passing through a  Cauchy surface of a conserved bounded zero or causal future-directed $C^1$-current is the same for all these surfaces (\ref{GMTA}).
\\
\hspace*{6mm}
{This result} of (\ref{GMTA}) is extended to all maximal achronal sets {containing the origin} under the assumption  on the decay (\ref{ALCCC})(b) of the probability current. Due to this decay the extension is achieved simply by flattening the maximal achronal sets becoming $\gamma$-achronal  (see \ref{CSCS} lemma) for $\gamma<1$.
The assumed decay is determined by the free relativistic time evolution of a massive particle. Regarding  the application to the massive scalar boson this assumption turns out to be a technicality satisfied for all relevant currents.
\\
\hspace*{6mm}
\textbf{Construction of achronal localization.} At this juncture one is ready to derive the main result (\ref{ALCCC}). Roughly speaking, every covariant conserved $C^1$ current  with bounded zeroth component being positive quadratic on Euclidean space determines by the flux through the achronal sets a covariant achronal localization.
\\
\hspace*{6mm}
\textbf{Application to the massive scalar boson.} 
The localizations of {the elementary particle being} the massive scalar boson in Euclidean space  (i.e., the  Euclidean covariant positive operator valued normalized measures in $\R^3$, called POL in \cite{C24})  are determined by the integrals of a positive definite rotational invariant probability density $J_0$ over the  regions of localization \cite[sec.\,6]{C24}. $J_0$ is the zeroth component of a covariant conserved current $\mathfrak{J}$ if and and only if the kernel is causal (\ref{CK}). This result is by \cite{GGP67}. Causal  kernels  have been studied in \cite{GGP67},\,\cite{HW71},\,\cite{C24}. Under the physically  irrelevant condition that the causal kernel is $C^4$  one verifies that $\mathfrak{J}$ satisfies the assumption in (\ref{ALCCC}) thus giving rise to a covariant achronal localization of the massive scalar boson.
\\
\hspace*{6mm}
Analogously one obtains a {covariant} family of  achronal localizations related to the  stress energy tensor of the massive scalar boson \cite{M23}.
\\
\hspace*{6mm}
\textbf{Representation of the causal logic.} Every (covariant) achronal localization determines uniquely a (covariant) representation of the causal logic and vice versa \cite[(21)]{C24}. Obviously this one-to-one correspondence emphasizes further the relevance of achronal localization. 
 \\
\hspace*{6mm}
Hence {apparently} for the first time a covariant representation of the causal logic for an elementary relativistic quantum mechanical system  is achieved. Also a covariant family of representations of the causal logic  is derived corresponding  to the above mentioned family of achronal localizations related to the stress-energy tensor.

\section{Notations and notions} \label{NN}

{\textbf{Minkowski spacetime.}} Vectors in $\R^4$ are denoted  by  $\mathfrak{x}=(x_0,x)$ with $x:=(x_1,x_2,x_3)\in\R^3$. Let $\varpi:\R^4\to\R^3$ denote the projection $\varpi(\mathfrak{x}):=x$.  Representing Minkowski spacetime by $\R^4$ the Minkowski product of $\mathfrak{a},\mathfrak{a}'\in\R^4$ is given by $\mathfrak{a}\cdot \mathfrak{a}':=a_0a_0'-aa'$, where for vectors $a,a'$ in $\R^3$ the scalar product $a_1a'_1+a_2a'_2+a_3a'_3$ is denoted by $aa'$. Often we use the notation $\mathfrak{a}^{\cdot 2}:=\mathfrak{a}\cdot\mathfrak{a}$. 
\\
{\textbf{Poincar\'e group.}} $\tilde{\mathcal{P}}=ISL(2,\C)$  is the universal covering group 
of  the {proper orthocronous} Poincar\'e group.    It acts on $\R^4$ as
\begin{equation}\label{PTUCH} 
g\cdot \mathfrak{x}:=\mathfrak{a}+\varLambda(A) \mathfrak{x}\quad  \text{ for } g=(\mathfrak{a},A)\in\tilde{\mathcal{P}}, \, \mathfrak{x}\in \R^4
\end{equation}
where $\varLambda:SL(2,\C)\to O(1,3)_0$ is  the universal covering homomorphism onto the proper orthochronous Lorentz group. For short one writes  $A\equiv (0,A)$, 
$\mathfrak{a} \equiv (\mathfrak{a},I_2)$, and $A\cdot \mathfrak{x}=\varLambda(A)\mathfrak{x}$. 
For $M\subset\R^4$ and $g\in \tilde{\mathcal{P}}$ define $g\cdot M:=\{g\cdot \mathfrak{x}: \mathfrak{x}\in M\}$.\\
\hspace*{6mm}
The group operation on  $\tilde{\mathcal{P}}$ reads 
$(\mathfrak{a},A)(\mathfrak{a}',A')
=(\mathfrak{a}+A\cdot\mathfrak{a}',AA')$ with identity element $(0,I_2)$ and inverse  $(\mathfrak{a},A)^{-1}=(-A^{-1}\cdot \mathfrak{a},A^{-1})$.
\\
{\textbf{Spacetime relations.}}  The fourvector $\mathfrak{z}\in\R^4\setminus\{0\}$ is  called \textbf{timelike, lightlike, causal} if $|z_0|>|z|$, $|z_0|=|z|$, $|z_0|\ge |z|$, respectively. It is \textbf{future-directed} if $z_0>0$.
\\
\hspace*{6mm}
The set $\mathfrak{a}+\R\mathfrak{z}$ for  $\mathfrak{a}, \mathfrak{z}\in\R^4$, $\mathfrak{z}\ne 0$ is called a {\bf line}. The line is {\bf timelike, lightlike, causal},  if so is $\mathfrak{z}$.
 \\
\hspace*{6mm}
A set  $\Delta\subset \R^4$ is said to be \textbf{spacelike} if {different points of $\Delta$ are spacelike separated, i.e.,} $|x_0-y_0| < |x-y|$ for $\mathfrak{x},\mathfrak{y} \in \Delta$,  $\mathfrak{x}\ne \mathfrak{y}$.
 \\
{\textbf{Achronal sets.}}
A set $\Delta\subset \R^4$ is said to be \textbf{achronal} if {the points of $\Delta$ are achronal separated, i.e.,} $|x_0-y_0| \le |x-y|$ for $\mathfrak{x},\mathfrak{y} \in \Delta$. 
 \\
\hspace*{6mm}
{Clearly every spacelike set is achronal.}
By definition $\Delta$ is {\bf maximal achronal} if $\Delta$ is not properly contained in an achronal set. An achronal set  is maximal achronal if and only if it meets every timelike  line. 
 \\
\hspace*{6mm}
{In Minkowski space, {\em Cauchy surfaces} \cite{O83}  are maximal achronal sets, which meet all lightlike lines. The converse holds also true \cite[ Moretti (36)]{C24}. The mass shell is an example of a spacelike maximal achronal set, which is not a Cauchy surface. Being maximal it is not  even contained in a Cauchy surface.} 
\\
\hspace*{6mm}
Note also  that {every  achronal set is contained in a maximal one and that} a maximal achronal set is always closed.
 \\
\hspace*{6mm}
{Most important is the fact that a set $\Delta\subset \R^4$ is achronal if and only if it is the graph of a \textbf{$1$-Lipschitz function} $\tau:\varpi(\Delta)\to \R$. It is maximal achronal iff $\varpi(\Delta)=\R^3$.}
 \\
\hspace*{6mm}
{Let the maximal achronal set $\Lambda=\{\big(\tau(x),x\big):x\in\R^3\}$  be endowed with the induced topology from $\R^4$.   Then $j:\R^3\to \Lambda$, $j(x):=(\tau(x),x)$  is a homeomorphism, $\{\mathfrak{x}\in\R^4: x_0<\tau(x)\}$ and  $\{\mathfrak{x}\in\R^4: x_0>\tau(x)\}$ are open with  boundary $\Lambda$, and  $\Lambda$ is path-connected. In particular  $\Lambda$ is a topological hypersurface. Clearly $\Lambda\subset \R^4$ has no interior points.}
 \\
\hspace*{6mm}
{ For more details on achronal sets see \cite[ sec.\,2.2]{C24}.}

\section{Divergence theorem on open sets with almost Lipschitz boundary}\label{DTALB}

{The main result  of this section is (\ref{MT}) Theorem. It is a version of the well-known Divergence Theorem for an open bounded subset of $\R^n$ with a rather weak regularity of its boundary.}
The following is inspired by Maggi \cite[Remark 9.5, Theorem 9.6]{M12}. {Repeatedly one refers to the very well-known {\em Area Formula}  in geometric measure theory first proved by H. Federer, see \cite[3.2., in particular 3.2.20]{F69}, citing  the easier accessible \cite[Theorem 9.1]{M12}. Part (a) of the proof of the preparatory result (\ref{GGL})   is the standard step familiar from the proof of the classical Gau\ss 's theorem. See e.g. \cite[ proof of (A8-16)]{A12} or the more elementary Step one  of the proof of \cite[Theorem 9.3]{M12}. }

As to the notation,   for $x\in\R^n$, $z\in\R^{n-1}$,  $s>0$  put
 $x':=(x_1,\dots,x_{n-1})$ and $x=(x',x_n)$, and  let $B_s(x):=\{y\in\R^n:|y-x|<s\}$,
$$C(x,s):=\{y\in\R^n: |y'-x'|<s,|y_n-x_n|<s\}\:, D(z,s):=\{y\in\R^{n-1}: |y-z|<s\}\:.$$  Given an outer measure $\omega$ on $\R^n$ and $X\subset \R^n$, the trace (restriction)  of $\omega$ on $X$ is the outer measure $\omega|X$ on $\R^n$ given by $\omega|X(F):=\omega(F\cap X)$. $\mathcal{H}^{n-1}$ denotes the $n-1$-dimensional Hausdorff measure on $\R^n$.

\begin{Def}\label{ALB} Let $E\subset \R^n$ be open. $E$ has \textbf{almost Lipschitz boundary}  $\partial E$ if there is a \textbf{localization} of $\partial E$ as follows. There is
a closed set $M_0\subset \partial E$ with $\mathcal{H}^{n-1}(M_0)=0$ such that
 for every $x\in M:=\partial E\setminus M_0$ there exists $s\in]0,\infty[$ with, up to rotation\footnote{This means that there exists an orthonormal basis $e_1,\dots,e_n$ of $\R^n$
 such that (a),(b) hold for the coordinates of $x$ with respect to this basis.},
 \begin{itemize}
\item[(a)] $C(x,s)\cap E=\{y\in C(x,s): y_n>u(y')\}$
\item[(b)]  $C(x,s)\cap \partial E=C(x,s)\cap M= \{y\in C(x,s): y_n=u(y')\}$
\end{itemize}
 for some Lipschitz function $u:D(x',s)\to \R$.
 {Let $G$ be  the set of points where $u$ is differentiable and}
  define  the unit vector field
 \begin{equation*}
 \nu:\operatorname{graph}u|_G\to \R^n, \quad     \nu(z,u(z)):=\frac{(\nabla u(z),-1)}{|(\nabla u(z),-1)|}
 \end{equation*}
Recall that by {\em Rademacher's theorem} $G$ is the complement of a Lebesgue null set and that $\nabla u$ is measurable. 

\end{Def}

Henceforth  $E\subset \R^n$ is open with almost Lipschitz boundary. The notation refers to (\ref{ALB}).

\begin{Lem}\label{RMS}   $\mathcal{H}^{n-1}|C(x,s)\cap M$ is a Radon measure and $$\mathcal{H}^{n-1}\big((C(x,s)\cap M)\setminus  \operatorname{graph}u|_G \big)=0\:.$$ 
\end{Lem}\\
{\it Proof.} 
Put  $D:=D(x',s)\cap \{z\in\R^{n-1}: |u(z)-x_n| <s\}$. Note   $ \operatorname{graph}\, u|_D =C(x,s)\cap \operatorname{graph}\, u= C(x,s)\cap \partial E = C(x,s)\cap M$. 
\\
\hspace*{6mm}
  $\mathcal{H}^{n-1}|C(x,s)\cap M=\mathcal{H}^{n-1}|\operatorname{graph}\,u|_D$ is a Radon measure due to the  area formula  \cite[Theorem 9.1]{M12} . Moreover, one has
$$\mathcal{H}^{n-1}(\operatorname{graph}\, u|_{D(x',s)\setminus G})=\int_{D(x',s)\setminus G}|(\nabla u(z),-1)|\d z =0\:.$$ \qed

The following preparatory result  (\ref{GGL}) comprises by the case $M_0=\emptyset$  the divergence theorem on bounded open sets with Lipschitz boundary  (see also e.g. \cite[Remark 9.5]{M12} and \cite[A8.8]{A12}).

\begin{Pro}\label{GGL} Let $E$ be bounded. Let  $\varphi \in C_c^1(\R^n)$ vanish on a neighborhood of $M_0$. Then
$$\int_E\nabla\varphi\; \d \mathcal{L}^n=\int_{\partial E}\varphi\, \nu_E\;\d \mathcal{H}^{n-1}$$
holds. Here $\nu_E$ is a $\mathcal{H}^{n-1}$-a.e. determined unit vector field on $\partial E$.  Locally $\nu_E$ coincides with $\nu$ in \emph{(\ref{ALB})}.
\end{Pro}
\\
{\it Proof.} (a)  Assume  first $\varphi \in C_c^1(C(x,s))$. By (\ref{RMS}) the surface integral is well-defined. Following step one of the proof of \cite[Theorem 9.3]{M12} the result follows due to $C(x,s)\cap \partial E=C(x,s)\cap M$ by definition (\ref{ALB})(b). Note that the weak gradient $\nabla f_\delta$ equals  $\frac{1}{2\delta}(-\nabla u(z),1)$ at  $z\in G$.
\\
\hspace*{6mm}
(b) Now the vector field $\nu_E$ is constructed. Applying the result in (a) to  all $\varphi \in C_c^1(C(x,s)\cap C(\bar{x},\bar{s}))$ shows that $\nu$ and $\bar{\nu}$ coincide a.e. on their common domain. There are countably many $C(x_m,s_m)$ from the localization in  (\ref{ALB}), which cover $M$. (Indeed, $M_0$ is closed and hence $G_\delta$.  Since $\partial E$ is compact it follows that $M=\partial E\setminus M_0$ is $\sigma$-compact.) Thus we may compose from the corresponding  $\nu_m$ by means of \cite[Theorem 9.1]{M12} an $\mathcal{H}^{n-1}$-a.e. determined unit vector field $\nu_E$ on $\partial E$ which locally coincides   a.e. with $\nu$ from  (\ref{ALB}).
\\
\hspace*{6mm}
(c) As to the general case let $A\supset M_0$ be open with $\varphi|_A=0$.  Note $\overline{E}\setminus A = (E\cup M)\setminus A \subset E\cup \bigcup_{x\in M}C(x,s)$. Since  $\overline{E}\setminus A $ is compact it is covered by $E$ and  finitely many $C(x,s)$. Hence one obtains a finite open cover of $\overline{E}$ by  $U_0:=E$, $U_k$ being some $C(x,s)$ for $k=1,\dots, N$, and $U_{N+1}:=A$.
\\
\hspace*{6mm}
Let $(\eta_k)_{k=0,\dots,N+1}$ be a partition of unity for $\overline E$ subject to this cover \cite[4.20]{A12}, i.e. $\eta_k\in C^\infty_c(\R^n)$, $\operatorname{supp}\eta_k\subset U_k$, $\eta_k\ge 0$, and $\sum_{k=0}^{N+1}\eta_k(x)=1$ for $x\in\overline{E}$. Then 

\begin{itemize}
\item $\int_E\nabla(\eta_0\varphi)\d \mathcal{L}^n=0$ by the fundamental theorem of calculus,  and obviously  $\int_{\partial E}(\eta_0\varphi) \nu_E\d \mathcal{H}^{n-1}=0$   

\item $1\le k\le N$: $\int_E\nabla(\eta_k\varphi)\d \mathcal{L}^n=\int_{\partial E}(\eta_k\varphi) \nu_E\d \mathcal{H}^{n-1}$  by (a),\,(b)

\item  $ k= N+1$: $\int_E\nabla(\eta_{N+1}\varphi)\d \mathcal{L}^n=0$ and $\int_{\partial E}(\eta_{N+1}\varphi) \nu_E\d \mathcal{H}^{n-1}=0$ as $\varphi|_A=0$

\end{itemize}
whence  $$\int_E\nabla \varphi \d \mathcal{L}^n=\sum_{k=1}^N\int_E\nabla(\eta_k\varphi)\d \mathcal{L}^n= \sum_{k=1}^N\int_{\partial E}(\eta_k\varphi) \nu_E\d \mathcal{H}^{n-1}=
\int_{\partial E} \varphi \nu_E \d \mathcal{H}^{n-1}\:.$$
\\
\hspace*{6mm}
(d) In conclusion consider a further localization of $\partial E$ according (\ref{ALB}) with $\tilde{\nu}_E$ the related unit vector field by (b). Then  $\int_{\partial E} \varphi \, \nu_E \;\d \mathcal{H}^{n-1}= \int_{\partial E}\varphi\, \tilde{\nu}_E\;\d \mathcal{H}^{n-1}$ holds for all $\varphi \in C_c^1(\R^n)$ with $\operatorname{supp}\varphi\cap M_0=\emptyset$, whence $\nu_E=\tilde{\nu}_E$  $\mathcal{H}^{n-1}$-a.e.\qed

\begin{Lem}\label{BRSM}  Let $X\subset \R^n$ be Borel. Then  $\mathcal{H}^{n-1}|X$ is Borel regular.
\end{Lem}
\\
{\it Proof.} Let $F\subset \R^n$.  Since $\mathcal{H}^{n-1}$ is Borel regular, there are Borel sets $A, B\subset \R^n$ with $F\cap X\subset A$, $F\setminus X\subset B$ and  
 $\mathcal{H}^{n-1}(F\cap X)=\mathcal{H}^{n-1}(A)$,  $\mathcal{H}^{n-1}(F\setminus X)=\mathcal{H}^{n-1}(B)$. Then the Borel set $C:=(A\cap X) \cup (B\setminus X)$ satisfies $F\subset C$ with $F\cap X= A\cap X= C\cap X$ so that $\mathcal{H}^{n-1}(F\cap X)=\mathcal{H}^{n-1}(C\cap X)$, i.e.,  $\mathcal{H}^{n-1}|X(F)=\mathcal{H}^{n-1}|X (C)$. \qed\\
 
{We are in a position to state and prove  our version  (\ref{MT})  of the Divergence Theorem. It distinguishes itself in the weakest  hypotheses concerning  $\partial E$ regularity we know. Thus (\ref{MT}) applies to the proof of the crucial technical result  (\ref{GMTA})}.

\begin{The}\label{MT}  Let $E$ be open bounded with almost Lipschitz boundary. Assume  $\mathcal{H}^{n-1}(\partial E)<\infty$  and  the Minkowski content $\mathcal{M}^{n-1}(M_0)=0$.  Then   $\mathcal{H}^{n-1}|\partial E$   is a Radon measure and for $\varphi \in C_c^1(\R^n)$
$$\int_E\nabla\varphi\; \d \mathcal{L}^n=\int_{\partial E}\varphi\, \nu_E\;\d \mathcal{H}^{n-1}$$
holds. Here $E$ can be replaced by $\overline{E}$ as $\mathcal{L}^n(\partial E)=0$, and  $\partial E$ can be replaced by $M$ as  $\mathcal{H}^{n-1}(M_0)=0$. 
\end{The}\\
{\it Proof.} $\mathcal{H}^{n-1}|\partial E$   is a Radon measure  by (\ref{BRSM}).   
Let $\delta>0$ and let $A_\delta$ denote  the $\delta$-neighborhood of $M_0$. Then
  \begin{itemize}
\item $\mathcal{H}^{n-1}(\partial E\cap A_\delta)\to 0$ for $\delta\to 0$
\item   $\delta^{-1}\mathcal{L}^n(A_\delta)\to 0 $ for $\delta \to 0$; in particular, $\mathcal{L}^n(A_\delta)\to 0$
\end{itemize}
Indeed, the first claim holds as  $\mathcal{H}^{n-1}(\partial E\cap A_\delta)<\infty$ and $A_\delta\downarrow_\delta M_0$. The very definition of the Minkowski content \cite[3.2.27]{F69} implies the second claim.
 \\
\hspace*{6mm} 
Furthermore by \cite[4.19]{A12} there is $f\in C^\infty_c(\R^n)$ with $\operatorname{supp}f\subset A_\delta$, $0\le f\le 1$, $f|_{M_0}=1$, and $|\nabla f|\le C/\delta$, where the finite constant $C$ does not depend on $\delta$.  
\\
\hspace*{6mm} 
Note that (\ref{GGL}) applies to $(1-f)\varphi$. So $\int_E\nabla \varphi \d \mathcal{L}^n=\int_{\partial E}(1-f)\varphi\, \nu_E\;\d \mathcal{H}^{n-1}+\int_E\nabla (f\varphi) \d \mathcal{L}^n=\int_{\partial E}\varphi\, \nu_E\;\d \mathcal{H}^{n-1}-\int_{\partial E}f\varphi\, \nu_E\;\d \mathcal{H}^{n-1}+\int_E(\nabla f)\varphi \d \mathcal{L}^n+
\int_E f \nabla\varphi \d \mathcal{L}^n$. The last three summands vanish as $\delta\to 0$.
 \\
\hspace*{6mm}
Indeed, $|\int_{\partial E}f\varphi\, \nu_E\;\d \mathcal{H}^{n-1}|\le   \int_{\partial E} 1_{A_\delta} |\varphi |  \d \mathcal{H}^{n-1}\le ||\varphi||_\infty \mathcal{H}^{n-1}(\partial E\cap A_\delta)\to 0$. Next, the crucial one,  $|\int_E(\nabla f)\varphi \d \mathcal{L}^n|\le \int_E C\delta^{-1} 1_{A_\delta} |\varphi| \d \mathcal{L}^n= ||\varphi||_\infty C \delta^{-1} \mathcal{L}^n(A_\delta)\to 0$. Finally one has $|\int_E f \nabla\varphi \d \mathcal{L}^n| \le\int_E1_{A_\delta} ||\nabla \varphi|| \d \mathcal{L}^n
\le ||\nabla\varphi||_\infty  \mathcal{L}^n(A_\delta)\to 0$, thus accomplishing the proof.\qed\\

An obviously equivalent formulation of (\ref{MT}) is

 \begin{Cor}\label{MTD} Let $E$ be as in  \emph{(\ref{MT})}   and $\mathfrak{v}\in C_c^1(\R^n,\R^n)$ be a vector field. Then
$$\int_E \operatorname{div}\, \mathfrak{v}\; \d \mathcal{L}^n=\int_{\partial E}\mathfrak{v}\, \nu_E\;\d \mathcal{H}^{n-1}$$
where $\operatorname{div}\,\mathfrak{v} := \sum_{i=1}^n\partial_iv_i$ denotes the divergence of $\mathfrak{v}$ and $\mathfrak{v}\, \nu_E$ is the $\R^n$-scalar product of $\mathfrak{v}$ and $\nu_E$.
\end{Cor}

In view of an application of (\ref{MT}),\,(\ref{MTD}) the following remarks on $\mathcal{H}^{n-1}(\partial E)<\infty$ and  $\mathcal{M}^{n-1}(M_0)=0$ may be useful. Note first that  $\mathcal{M}^{n-1}(M_0)=0$ implies   $\mathcal{H}^{n-1}(M_0)=0$.
\\
\hspace*{6mm}
According to  \cite[Theorem 8.1]{M12},  $\mathcal{H}^{n-1}(D)<\infty$ if $D$ is the image  under an injective Lipschitz function on $\R^{n-1}$ in $\R^n$ of a  Lebesgue measurable set  $C$ of finite Lebesgue measure.   If $C$ is a Lebesgue null set then $\mathcal{H}^{n-1}(D)=0$.

\begin{Cor} Let $\partial E$ be covered by finitely many sets $D_i$, each $D_i$ being the image under an injective Lipschitz function on $\R^{n-1}$ in $\R^n$ of a  Lebesgue measurable $C_i$ of finite Lebesgue measure. Then $\mathcal{H}^{n-1}(\partial E)<\infty$.

 If the images $D_i$ of countably many Lebesgue null sets $C_i$ cover $M_0$, then  $$\mathcal{H}^{n-1}(M_0)=0\:.$$
\end{Cor}
\\
\hspace*{6mm}
According to \cite[Theorem 3.2.39]{F69}, $\mathcal{H}^{n-1}(D)=\mathcal{M}^{n-1}(D)=0$ holds, if  $D\subset \R^n$ is closed and if there is a Lipschitz function $h:\R^{n-1}\to \R^n$ mapping some bounded  $C\subset \R^{n-1}$ onto $D$
or, equivalently, if $h$ is locally Lipschitz defined on $\overline{C}$ with $h(C)=D$. The latter holds because $\overline{C}$  is compact.

\begin{Pro}\label{MGOA} Let $M_0$ be the union of finitely many sets $D_i$ of Minkowski content $\mathcal{M}^{n-1}(D_i) =0$. Then  $\mathcal{M}^{n-1}(M_0) =0$. 

 Moreover, $\mathcal{M}^{n-1}(D_i) =0$ holds if $\mathcal{H}^{n-1}(D_i) =0$ and if $D_i$ is the image of a compact set under a locally Lipschitz function.
\end{Pro}\\
{\itshape Proof.} It remains to prove the first part of the assertion. The finite subadditivity of the upper Minkowski content yields $\mathcal{M}^{*n-1}(M_0)       \le \sum_i\mathcal{M}^{*n-1}(D_i)  = \sum_i\mathcal{M}^{n-1}(D_i)=0$ implying    $\mathcal{M}^{n-1}(M_0) =0$.        \qed

\section{Flux passing through a maximal achronal set}

{The main result  of this section is (\ref{GMTA}) Theorem, v.i.} {It extends \cite[Proposition 37]{DM24}  from smooth Cauchy  surfaces to all maximal achronal sets in Minkowski spacetime. Note that \cite[Proposition 37]{DM24} is equally valid  for a $C^1$ vector field in place of a smooth one. 
\textbf{Cauchy surfaces}, studied in
general spacetime  theories,  are the sets which meet every inextendible timelike smooth curve 
exactly once \cite[Chapter 14, Definition 28]{O83}.  By \cite[Chapter 14, Lemma 29]{O83} they even meet all inextendible causal smooth curves.
According to   \cite[(9) Remark, (36) Theorem]{C24} a    Cauchy surface {in Minkowski spacetime} is just 
a maximal achronal   set which  intersects every lightlike line.
A maximal achronal set equals the graph of the corresponding $1$-Lipschitz function with domain $\R^3$. }
\\
\hspace*{6mm}
{The basic idea of the proof of the crucial technical result  (\ref{GMTA}) is the application of the Divergence Theorem to a distorted cylindrical surface in $\R^4$ (which is   $\partial E_n$ in the proof of   (\ref{GMTA})) with a flat bottom base, an achronal top base, and the side constituted by  integral curves of the divergence-free $C^1$ vector field. As the flux through the side is zero, the fluxes through the bases must be equal. This is the desired result. The crux is that the versions of the Divergence Theorem known to us do not apply to this kind of surface, which obviously is not polyhedral, nor $C^1$ as the top base need not be $C^1$.  Actually, it is not even Lipschitz, but almost Lipschitz. In the sense of definition  (\ref{ALB}) the set $M$ consists of the  interior points of the side and the bases. The complement $M_0$ of $M$ is the union of the boundaries of the bases. Here the extension (\ref{MTD}) of the Divergence Theorem applies as the exceptional set $M_0$ is small enough, i.e. its $3$-dimensional Minkowski content is zero.}

\begin{Lem}\label{CSCS} Let $\tau:\R^3\to\R$ and  $S:=\{(\tau(x),x): x\in\R^3\}$. Then $S$ is a spacelike  (according to \emph{sec.\,1}) Cauchy surface if $|\tau(x)-\tau(y)|<|x-y|$ for $x\ne y$ and  $\limsup_{|x|\to\infty}|\tau(x)| / |x|<1$ or if a fortiori $\tau$ is  $L$-Lipschitz with $L<1$, {in which case $S$ is called $L$-achronal.} 
\end{Lem}
\\
{\it Proof.} Obviously $S$ is spacelike. Assume  $(\mathfrak{a}+\R\mathfrak{z})\cap S=\emptyset$ for some $\mathfrak{z}=(1,e)$ with $0<|e|\le 1$. Let $a_0>\tau(a)$. (The case $a_0<\tau(a)$ is analogous.) Then  by continuity $a_0+s>\tau(a+se)$ for all $s$.
Hence for $s<-a_0$ one has $ |\tau(a+se)|  /|a+se|>|a_0+s|/|a+se|\to 1/|e|\ge 1$ for $s\to -\infty$. This contradicts the assumption $\limsup_{|x|\to\infty}|\tau(x)| / |x|<1$.\qed\\

A $C^1$ vector field $\mathfrak{j}$ on $\R^4$ is said to satisfy the \textbf{continuity equation} if $\operatorname{div}\mathfrak{j}=0$, i.e., if 
\begin{equation}\label{CERCO}
 \partial_0 j_0(\mathfrak{x}) +\partial_1j_1(\mathfrak{x})   +\,\partial_2j_2(\mathfrak{x})+ \partial_3 j_3(\mathfrak{x})   =0
 \end{equation}
holds  for  $\mathfrak{x}\in\R^4$. 
 With standard relativistic notation (\ref{CERCO}) reads $\partial_\mu J^\mu=0$.

{\begin{The}\label{GMTA}
Let the  real $C^1$ vector field $\mathfrak{j}$ on $\R^4$  be bounded and  satisfy the continuity equation \emph{(\ref{CERCO})}. Suppose that $\mathfrak{j}$ is zero or causal future-directed,  i.e.
\begin{itemize}
\item $j_0(\mathfrak{x})\ge |j(\mathfrak{x})|$  for all $\mathfrak{x}\in\R^4$
\end{itemize}
Then for every maximal achronal set  $\Lambda$ being the graph of the corresponding $1$-Lipschitz function  $\tau: \R^3\to\R$,  the future-directed  {\bf flux} of $\mathfrak{j}$ passing through  $\Lambda$ (the left-hand side of the inequality below) satisfies
\begin{equation}\label{ITAU}
\int \big( j_0(\tau(x),x) -  j(\tau(x),x)\nabla\tau(x)\big)  \d^3x  \le \int j_0(0,x) \d^3x \:.
\end{equation}
The integrands are nonnegative and the integrals may be infinite, $\nabla \tau$ is measurable a.e. determined. If $\Lambda$ is a Cauchy surface then in \emph{(\ref{ITAU})} equality holds.
\end{The}
\\
{\it Proof.} 
By the assumption on $\mathfrak{j}$ and $|\nabla \tau|\le1$ the integrands are nonnegative. 
Let $k:\R^3\to\R$ be positive bounded integrable $C^1$ like $k(x)=(1+x^2)^{-2}$.
\\
\hspace*{6mm}
 Then the $C^1$ vector field $\mathfrak{v}=(v_0,v):=(j_0+k,j)$ is bounded  so that its  flow is $C^1$, complete and global, satisfies the continuity equation, and  everywhere holds $v_0 > |v|$. Hence the integral curves $\gamma_\mathfrak{y}$ of $\mathfrak{v}$,  determined by $\dot{\gamma}_\mathfrak{y}(s)=\mathfrak{v}(\gamma_\mathfrak{y}(s))$, 
$\gamma_\mathfrak{y}(0)=\mathfrak{y}$, $\mathfrak{y}\in\R^4$, exist for all $s\in\R$, and are timelike  future directed. Furthermore $(\gamma_\mathfrak{y}(\cdot))_0$ is strictly increasing and $\gamma_\mathfrak{y}(\R)\cap \Lambda$ is empty or a singleton.
\\
\hspace*{6mm}
 Let $$U:=\{x\in\R^3: \gamma_{(0,x)}(\R)\cap\Lambda\ne \emptyset\}$$ Note that $U=\R^3$  if $\Lambda$ is a Cauchy surface.
 For $x\in U$ let $\sigma(x)\in \R$ be uniquely determined by $\gamma_{(0,x)}(\sigma(x))\in\Lambda$. Consider $$\mathfrak{h}:U\to \R^4,\; \mathfrak{h}(x):=\gamma_{(0,x)}(\sigma(x))$$ Note that $\mathfrak{h}$ is injective and $\mathfrak{h}(U)=\Lambda$. }
 \\
\hspace*{6mm}
(a) {Here it is shown that  $U$ is open,  $\sigma$  and $\mathfrak{h}$ are Lipschitz continuous, and $U$ and  $\Lambda$ are homeomorphic by $ \mathfrak{h}$.}
\\
\hspace*{6mm}
{Indeed, recall  \cite[Proposition 37]{DM24} by which $$\Phi:\R\times\R^3\to\R^4, \;\Phi(s,x):=\gamma_{(0,x)}(s)$$ is a diffeomorphism. Write $\Phi=(\varphi,\phi)$ and let $f:\R\times U\to\R$, $f:=(\varphi -\tau\circ\phi)$. Note that $f(s,x)=0$ holds if and only if $x\in U$ and  $s=\sigma(x)$.
 \\
\hspace*{6mm}
 If $\tau$ is $C^1$ then so is $f$ and, since $\Phi=(\varphi,\phi)$, $\Phi(s,x)=\gamma_{(0,x)}(s)$ and $\dot{\gamma}_{(0,x)}(s)=\mathfrak{v}(\gamma_{(0,x)}(s))$, one has 
\begin{equation}\label{PDAIF}
\partial_1f=v_0\circ\Phi -  ( \nabla\tau\circ\phi)\, (v\circ\Phi)\, \ge (v_0-|v|)\circ\Phi>0
\end{equation}
So  the implicit function theorem applies at any $x\in U$, whence  $\sigma$ is $C^1$ and hence locally Lipschitz. One turns to the general case.
\\
\hspace*{6mm}
Let $\tau$ be $1$-Lipschitz. Then by Rademacher's theorem there is $S\subset \R^3$ such that $\R^3\setminus S$ is a Lebesgue null set and $\tau$ is differentiable on $S$, i.e., 
\begin{equation}\label{DQ}
 \lim_{0\ne h\to 0}    \frac{1}{|h|} \big| \tau(x+h)-\tau(x)- \nabla\tau(x)h \big| =0 
 \end{equation}
 holds for every $x\in S$. Clearly $f$ is Lipschitz.  By the chain rule, $f$ is differentiable and $\partial_1f$ satisfies (\ref{PDAIF}) on $\Phi^{-1}(\R\times S)$. 
\\
\hspace*{6mm}
Let  $(r,a)\in\R\times\R^3$. Put $\alpha:= \frac{1}{2} (v_0-|v|)\circ\Phi(r,a)\,>0$. Then there is a neighborhood $V$ of  $(r,a)$ such that $(v_0-|v|)\circ\Phi(t,x)\ge \alpha$ for all $(t,x)\in V$.
Therefore $\partial_1f(x,t)\in [\alpha,\infty[$ for all $(x,t)\in V\cap \Phi^{-1}(\R\times S)$. According to the definition given in \cite[sec.\,2]{P77} this implies that the generalized derivative of $f(\cdot,a)$ at $r\in\R$ is contained in $ [\alpha,\infty[$ disjoint from $\{0\}$. Thus the Lipschitz implicit mapping theorem \cite[Theorem 12.1]{P77} applies. It follows that $U$ is open and that $\sigma$ is locally Lipschitz.
\\
\hspace*{6mm}
Since  $\mathfrak{h}(x)=\Phi(\sigma(x),x)$ holds  for $x\in U$, also $\mathfrak{h}$ is locally Lipschitz. Finally,  for every open $V\subset U$ one has $\mathfrak{h}(V)=\Phi(\R\times V) \cap \Lambda$, whence  $\mathfrak{h}(V)$ is open in $\Lambda$.}
\\
\hspace*{6mm}
(b)  {Here a distorted cylindrical surface $E_n\subset \R^4$, $n\in\N$, is defined to which the divergence theorem (\ref{MTD}) applies.} {First note that $q:U\to \R^3$, $q(x):=\varpi(\mathfrak{h}(x))$ is a locally Lipschitz homeomorphism.
\\
\hspace*{6mm}
 It is easy to construct  $C_n\subset U$ being the union of finitely many closed balls such that $q^{-1}(\{x\in\R^3:|x|\le n\})\subset C_n\subset C_{n+1}$. Obviously $\bigcup_nC_n=U$ and hence $\bigcup_n \mathfrak{h}(C_n)=\Lambda$.
Let $A_n$, $B_n$ denote the boundary and the interior of $C_n$. Now suppose  $\sigma(x)>0$ for $|x|<n$. Then 
$$E_n:=\bigcup_{x\in B_n}\{\gamma_{(0,x)}(s): 0< s < \sigma (x)\}$$
is open in $\R^4$ with $\partial E_n=M\cup M_0$, where $M:=\{0\}\times B_n \cup L_n \cup \mathfrak{h}(B_n)$ with $L_n:=\bigcup_{x\in A_n}\{\gamma_{(0,x)}(s): 0 < s < \sigma(x)\}$,  and $M_0:=\{0\}\times A_n\,\cup\, \mathfrak{h}(A_n)$. }
\\
\hspace*{6mm}
{Now the proof of the basic idea follows in (c)--(e). One verifies that {the divergence theorem} (\ref{MTD}) applies to $n=4$, $E =E_n$ showing $\mathcal{H}^3(\partial E_n)<\infty$ by (c),\,(d), and $\mathcal{M}^3(M_0)=0$ by (e),  (\ref{MGOA}). {It is decisive that the integral over $L_n$ is zero as $L_n$ is constituted of integral curves.}}
\\
\hspace*{6mm}
(c) {Let $D\subset\R^3$ be  bounded Borel. Then $\mathcal{H}^3(\{0\}\times D)<\infty$, $\mathcal{H}^3(\operatorname{graph} \tau|_D)<\infty$ by  \cite[Theorem 9.1]{M12}.
\\
\hspace*{6mm}
(d) Let $D$ be a bounded Borel subset of the tube $T_n:=\{\gamma_{(0,x)}(s):s\in\R, x\in A_n  \}$. The claim is  $\mathcal{H}^{n-1}(D)<\infty$. Note that $A_n$ is contained in the union of finitely many spheres. Hence it suffices to assume that $A_n$ is a sphere. Without restriction let $A_n=\{|x|=R\}$.
\\
\hspace*{6mm}
Then $A_n$ is covered by $A:=\{x\in A_n: |x_1|\le R/3,  |x_2|\le R/3\}$ and finitely many rotations of $A$ around the origin. Hence it suffices to show that $\mathcal{H}^{n-1}(D')<\infty$ for $D':=\{\gamma_{(0,x)}(s):|s|\le S, x\in A\}$ and $0<S<\infty$.
\\
\hspace*{6mm}
 Recall the diffeomorphism  $\Phi:\R\times \R^3\to \R^4$. Then $f:=\R\times ]-R/2,R/2[ \times$ $]-R/2,R/2[\to \R^4$, $f(s,x_1,x_2):=\Phi\big(s; x_1,x_2,\sqrt{R^2-x_1^2-x_2^2}\,\big)$ is injective $C^1$, whence  Lipschitz on compact sets, and $f([-S,S]  \times [-R/3,R/3] \times [-R/3,R/3])=D'$. The claim holds by  \cite[Theorem 8.1]{M12}. 
\\
\hspace*{6mm}
(e) Let $C\subset\R^3$ be a compact Lebesgue null set. Then $\mathcal{H}^{3}(\{0\}\times C)=0$ and $\mathcal{H}^{3}(\mathfrak{h}(C))=0$  hold by \cite[Theorem 8.1]{M12}. Indeed, the former is obvious, the latter holds true as $\mathfrak{h}$ is Lipschitz on $C$ being compact. It follows $\mathcal{M}^3(M_0)=0$ by (\ref{MGOA}).}
\\
\hspace*{6mm}
(f) {Here the Divergence Theorem is applied.} {Let $\eta\in C_c^1(\R^4)$ with $0\le \eta \le1$ and $\eta|_{\overline{E}_R}=1$. Put $\phi:=\eta \mathfrak{v}$.
  Then  (\ref{MTD})  yields $0=\int_M \mathfrak{v}\,\nu_{E_R} \d \mathcal{H}^{3}$ as $\mathfrak{v}$ satisfies the continuity equation. Moreover the integration over $L_R$ yields $0$ as the integrand is $0$. So
using \cite[Theorem 9.1]{M12} it follows  $\int_{\varpi(\mathfrak{h}(B_n))} \big(v_0(\tau(x),x) - v(\tau(x),x)\nabla \tau(x)\big) \d x^3 =\int_{B_n} v_0(0,x) \d x^3$.
\\
\hspace*{6mm}
(g) Finally, as explained in  \cite[Proposition 37]{DM24},  the condition $\sigma (x)>0$ for $|x|<n$ in (b)  can be removed and the limit $n\to\infty$ carried out. Following \cite[Proposition 37]{DM24}, one ends up with the  formula 
$$\int_{\R^3} \big(j_0(\tau(x),x) - j(\tau(x),x)\nabla \tau(x)\big) \d x^3 =\int_U j_0(0,x) \d x^3\le \int_{\R^3} j_0(0,x) \d x^3$$
with $U=\R^3$ if $\Lambda$ is a Cauchy surface.}
\qed

 Under the assumptions made on the vector field $\mathfrak{j}$, the result in (\ref{GMTA})  states that the future-directed flux passing through a  Cauchy surface  is the same for all these surfaces.
Under the additional assumption  (\ref{CMTA})(b) on $\mathfrak{j}$,    {this holds true for} all maximal achronal sets containing the origin {simply by flatten them becoming $\gamma$-achronal for $\gamma<1$}.
\hspace*{6mm}

\begin{Lem}\label{CMTA}
Let the real $C^1$ vector field $\mathfrak{j}$ be bounded and satisfy the continuity equation  \emph{(\ref{CERCO})}. Suppose that 
\begin{itemize}
\item[\emph{(a)}]    $j_0(\mathfrak{x})\ge |j(\mathfrak{x})|$ for  $\mathfrak{x}\in\R^4$
\item[\emph{(b)}] 
$j_0(\mathfrak{x})\le C(1+|x|)^{-N}$ for $ |x|\ge|x_0|$ with some constants $N>3$ and  $C<\infty$   
\end{itemize}
Then for every maximal achronal set  $\Lambda$ with $0\in\Lambda$
\begin{equation}\label{ETAU}
\int \big( j_0(\tau(x),x) -  j(\tau(x),x)\nabla\tau(x)\big)  \d^3x  = \int j_0(0,x) \d^3x 
\end{equation}
holds.  Here $\operatorname{graph}\tau=\Lambda$ with $\tau(0)=0$.
The integrands are nonnegative and the integrals are finite. 
\end{Lem}

{\itshape Proof.} Due to (a) and $|\nabla \tau|\le1$ the integrands are nonnegative. Note $|\tau(x)|=|\tau(x)-\tau(0)|\le |x-0|=|x|$. Hence, by (b),  $ \big| j_0(\tau(x),x) -  j(\tau(x),x)\nabla\tau(x)\big|\le 2 j_0({\tau(x)},x) \le 2C_N(1+|x|)^{-N}$ with $N>3$, whence
the integrals are finite.
\\
\hspace*{6mm}
Let $0<\gamma<1$. Then $\mathfrak{j}(\gamma x_0,x)\to \mathfrak{j}(x_0,x)$ for $\gamma\to 1$ by continuity.  Note that $\gamma \tau$ is $\gamma$-Lipschitz. Hence by 
(\ref{CSCS})  the corresponding maximal achronal set is a (spacelike) Cauchy surface. Therefore by (\ref{GMTA}) equation (\ref{ETAU}) holds for $\gamma \tau$ in place of $\tau$. Note that still $|\gamma \tau (x)|\le |x|$. Thus, by (b),  the map $\R^3 \ni x\mapsto 2C_N(1+|x|)^{-N}$ is an integrable majorant uniform with respect to $\gamma$, whence the claim by dominated convergence.\qed

\section{Covariant achronal localization out of covariant conserved current}\label{SALCCC}

 Let $\mathcal{H}$ be a separable Hilbert space.  Let $\mathcal{B}^{ach}$ denote the family of Borel subsets $\Delta$ of $\R^4$, which are achronal. 
 
\begin{Def}\label{POB} Let $T(\Delta)$ for $\Delta\in\mathcal{B}^{ach}$ be a nonnegative bounded operator on $\mathcal{H}$. Suppose  $T(\emptyset)=0$ and  $\sum_nT(\Delta_n)=I$   for every sequence $(\Delta_n)$ of mutually disjoint sets in $\mathcal{B}^{ach}$ such that $\bigcup_n\Delta_n$ is maximal achronal. Then the map $T$ is called 
an \textbf{achronal localization} (AL).
\\
\hspace*{6mm}
Let $W$ be a unitary representation of $\tilde{\mathcal{P}}$. Then the AL $T$ is said to be (Poincar\'e)  \textbf{covariant} by means of $W$ if $T(g\cdot\Delta)=W(g)T(\Delta)W(g)^{-1}$ holds for $g\in \tilde{\mathcal{P}}$ and $\Delta\in\mathcal{B}^{ach}$.
\end{Def}
\\
\hspace*{6mm}
 As mentioned the meaning of $T$ is that $\langle \phi,T(\Delta)\phi\rangle$ is the probability of localization of the quantum mechanical system  in the spacetime region $\Delta$ if the system is in the state represented by the unit vector $\phi\in  \mathcal{H}$. 
\\
\hspace*{6mm}
There exist AL whose the localization operators $T(\Delta)$ are orthogonal projections \cite[(22) Theorem]{C24}. In this case the  localization operators commute. However, a quantum mechanical system localized by a projection valued AL necessarily does   not have a semi-bounded  energy operator. This   no-go result following from Hegerfeldt's well-known  theorem \cite{Hegerfeldt}\footnote{{See also \cite{Hegerfeldt2} and D.Buchholz and J.Yngvason’s comment also on Hegerfeld's achievements \cite{BY}.}} 
regards a {\em first type} of Einstein causality requirement whose modern generalized  reformulation \cite{C17,C24} we shall  present below. 
\\
\hspace*{6mm}
The  notion of localization of the above definition seems appropriate  to describe measurement processes where {\em a quantum system is absorbed by the apparatus and no further localization  measurments can be performed on it}.  
This is because, in case of subsequent measurements a {\em second type} of  Einstein's causality requirement essentially regarding the  no-signaling condition has to be considered:
 {The statistics of the measurement results $\Delta$ must be the same, regardless of whether a non-selective measurement was made in $\Delta'$ or not.}

To comply with this requirement, under some popular assumptions about the post measurement state, $T(\Delta)$ and $T(\Delta')$ should commute if $\Delta$ and $\Delta'$ cannot be joined by causal curves.\footnote{{If we explicitly assume that $T(\Delta)$ and $T(\Delta')$ are projectors and suppose that the post-measurement state of a non-selective measurement of $T(\Delta')$ and $I-T(\Delta')$ is obtained with the Lüders-von Neumann projection postulate for every initial state, the no-signaling requirement easily implies that $T(\Delta)$ and $T(\Delta')$ commute. If  $T(\Delta)$ and $T(\Delta')$ are effects, the no-signaling condition is guaranteed if, more weakly, the respective Kraus operator commute. However, if} {we also assume, as is common in local quantum physics, that all relevant operators (including Kraus ones) are organized in local von Neumann algebras (here  associated to  the causal completions of $\Delta$ and $\Delta'$), the Kraus operator of $T(\Delta)$ must also commute with the {\em adjoint} of the  Kraus operator of $T(\Delta')$. In summary,  $T(\Delta)$ and $T(\Delta')$  commute.}} This requirement cannot be fulfilled as a consequence of {\em Malament's theorem} \cite{Malament} and its modern re-formulations {due to} {Halvorson and Clifton  particular \cite{HC}. A quick review on these  issues related to various facets of Einstein's causality and localization appears in the introduction of \cite{M23}.}
\\
\hspace*{6mm}
 The first-type of causality requirement  in the modern generalized  perspective states that the probability of localization in {{\em any region of influence}}  determined by the limiting velocity of light is not less than that in the region of actual localization. {More precisely, consider  a region $\Delta'$ which is contained in a causal base (namely a Cauchy surface which is a spacelike set). Then $\Delta'$ is a {\bf  region of influence} of the achronal region  $\Delta$} if all causal straight  lines, which intersect  $\Delta$, also meet $\Delta'$.
Hence the condition imposed by causality on an AL  reads 
\begin{equation*}
T(\Delta)\le T(\Delta')\:.  \tag{CC}
\end{equation*}
\\
{For more details see \cite{C24}.} 
\\
\hspace*{6mm}
In \cite{DM24},  condition CC was proved true in a special case where the sets $\Delta$ belong to smooth Cauchy surfaces and in \cite{C24}, the final result has been established that an AL satisfies CC  in full. As argued in  \cite{C24}, CC even necessitate achronal localization. In fact spacelike localization is not sufficient since CC induces the localization  in achronal hyperplanes. This fact is reported in \cite{C24} and studied in detail  in \cite{C25}. {An important result is that the concepts of achronal localization AL and representation of the causal logic RCL are equivalent (cf.(\ref{EALCLL})). {Furthermore,} just the existence of a maximal spacelike set, which is not maximal achronal because of a missing piece of an achronal not spacelike hyperplane
(see $P$ in \cite[(8) Example]{C24},)  is the reason behind the missing \textbf{orthomodularity} of the lattice of the causally complete sets generated by the spacelike relation (cf. \cite[sec.\,11.3]{C17}).} 
\\
\hspace*{6mm}
Hence we consider the study and the explicit construction of  the achronal localization of the massive scalar boson to be of utmost relevance. 
\\
\hspace*{6mm}
It is  the very principle of causality which  lets one think  of  the probability of localization as a conserved quantity reigned by an associated  density current $\mathfrak{J}$.
Indeed, by (\ref{ALCCC}) a covariant AL can be constructed  by means of a  covariant   conserved current assuming that on the Euclidean space its  zeroth component describes the density of the probability of localization, namely  explicitly $J_0(\phi;0,x)\ge 0$, $x\in\R^3$, and
\begin{equation}\label{BRLC}
\langle \phi,T(\Delta)\phi\rangle=\int_\Delta J_0(\phi;0,x) \d^3 x
\end{equation}
for Borel $\Delta\subset   \R^3$.   

\begin{Def}\label{CCC} {Let $W$ be a unitary representation of $\tilde{\mathcal{P}}$ and} let $\mathcal{D}$ be a $W$-invariant dense subspace of $\mathcal{H}$. Let  $\mathfrak{J}=(J_0,J)$ be a map from $\mathcal{D}\times \R^4$ to $\R^4$ such that $\mathfrak{J}(\phi,\cdot)$ is a bounded $C^1$ vector field. 

 (i) $\mathfrak{J}$ is a {\bf conserved current} if the latter satisfies  the continuity equation.

 (ii) $\mathfrak{J}$ is a (Poincar\'e) {\bf covariant current} if it holds  $$\mathfrak{J}\big(W(g)\phi,\mathfrak{x}\big)=A\cdot\mathfrak{J}\big(\phi,g^{-1}\cdot\mathfrak{x}\big)\:,\quad g=(\mathfrak{a},A)\in \tilde{\mathcal{P}}\:.$$
\end{Def}

The introduction of a suitable  dense space $\mathcal{D}$ in (\ref{CCC}) takes account of the fact that $\mathfrak{J}$ may be given, as in the case of the massive scalar boson, as an integral operator, which is not defined for all wave functions $\phi$. Moreover, $\mathfrak{J}$ is supposed to be real. Actually, by checking the proof of (\ref{FDP}), this is a consequence of $J_0$  being real and covariance.

\begin{Lem}\label{FDP}
 A covariant current satisfies $J_0(\phi,\mathfrak{x})\ge |J(\phi,\mathfrak{x})|$ for $\mathfrak{x}\in\R^4$, $\phi\in\mathcal{D}$, if and only if $J_0(\phi;0,x)\ge 0$ for $x\in\R^3$, $\phi\in\mathcal{D}$.
 \end{Lem}\\
{\itshape Proof.}  $J_0(\phi;0,x)\ge 0$ for $x\in\R^3$,  $\phi\in\mathcal{D}$ $\Leftrightarrow$ $\mathfrak{J}(W(g)^{-1}\phi;0,x)\cdot (1,0,0,0)\ge 0$ for $x\in\R^3$,  $\phi\in\mathcal{D}$,  $g=(\mathfrak{a},A)\in\tilde{\mathcal{P}}$
 $\Leftrightarrow$ $\mathfrak{J}(\phi; \mathfrak{a}+A\cdot(0,x))\cdot (A\cdot(1,0,0,0)\ge 0$ for $x\in\R^3$,  $\phi\in\mathcal{D}$,  $(\mathfrak{a},A)\in\tilde{\mathcal{P}}$  $\Leftrightarrow$ $\mathfrak{J}(\phi; \mathfrak{x})\cdot \mathfrak{e}\ge 0$ for $\mathfrak{x}\in\R^4$, $\phi\in\mathcal{D}$, $\mathfrak{e}^{\cdot2} =1$, $e_0>0$, whence the claim.\qed

 \begin{Def}\label{SQUA} A  map $q:\mathcal{D}\to \R$ is said to be \textbf{quadratic} if (i) $q(\lambda \phi)=|\lambda|^2q(\phi)$ for $\lambda\in\C$, $\phi\in\mathcal{D}$ and (ii) 
 $s_q(\phi,\phi'):=\frac{1}{4}\sum_{\zeta=1,-1,\i,-\i} \zeta q(\zeta \phi+\phi')$ for $\phi,\phi'\in\mathcal{D}$ is  Hermitian sesquilinear.  Note that $s_{q}(\phi,\phi)=q(\phi)$ holds by (i), whence (ii) is the polarization identity of $s_q$.
 \end{Def}

 \begin{Lem}\label{SPWQ} Assume \emph{(\ref{BRLC})} for $\phi\in\mathcal{D}$. Then, for every $x\in\R^3$,   $J_0(\cdot\, ;0,x)$ is quadratic.
 \end{Lem}
 \\
{\itshape Proof.}  Put   $q_x:=J_0(\cdot\, ;0,x)$.  Note  $\langle \phi,T(\Delta)\phi\rangle=\int_\Delta q_x(\phi)\d^3 x$ and hence  
$$\langle \phi,T(\Delta)\phi'\rangle=\int_\Delta s_{q_x}(\phi,\phi')\d^3 x\:,$$ where $\langle \phi,T(\Delta)\phi'\rangle$ is a Hermitian sesquilinear form. 
\\
\hspace*{6mm}
Hence, as to (\ref{SQUA})\,(i),  $\int_\Delta\big(q_x(\lambda \phi)-|\lambda|^2q_x(\phi)\big)\d^3x=0$ holds for every $\Delta$, whence the integrand is a.e. zero and by its continuity it is zero everywhere. 
\\
\hspace*{6mm}
Turn to  (\ref{SQUA})\,(ii). Show $s_{q_x}(\phi_1+\phi_2,\phi')-s_{q_x}(\phi_1,\phi')-s_{q_x}(\phi_2,\phi')=0$. Indeed, one has
 $\int_\Delta\big(s_{q_x}(\phi_1+\phi_2,\phi')-s_{q_x}(\phi_1,\phi')-s_{q_x}(\phi_2,\phi')\big)\d^3 x =0$ for every $\Delta$, whence the claim. The remaining properties  for $s_{q_x}$ regarding a Hermitian sesquilinear form hold analogously.\qed

\begin{Lem}\label{QDIAS} Let $\mathfrak{J}$ be covariant. Then $J_0(\cdot\, ,\mathfrak{x})-J(\cdot\, ,\mathfrak{x})\,e$ is quadratic on $\mathcal{D}$ for every $\mathfrak{x}\in\R^4$ and  $e\in\R^3$, $|e|\le 1$ if and only if $J_0(\cdot\, ;0,x)$ is quadratic on $\mathcal{D}$  for every $x\in\R^3$.
\end{Lem}
\\
{\itshape Proof.}  $J_0(\phi;0,x)$ is quadratic regarding $\phi$  for every  $x\in\R^3$ $\Leftrightarrow$ $J_0(W(g)^{-1}\phi;0,x)$  is quadratic regarding $\phi$  for every  $x\in\R^3$, $g=(\mathfrak{a},A)\in\tilde{\mathcal{P}}$ $\Leftrightarrow$ 
$\mathfrak{J}(W(g)^{-1}\phi;0,x)\cdot (1,0,0,0)$   = $\mathfrak{J}(\phi;\mathfrak{a}+A\cdot(0,x))\cdot (A\cdot(1,0,0,0))$   
is quadratic regarding $\phi$  for every  $x\in\R^3$, $(\mathfrak{a},A)\in\tilde{\mathcal{P}}$ 
  $\Leftrightarrow$ $\mathfrak{I}(\phi;\mathfrak{x})\cdot\mathfrak{e}$  is quadratic regarding $\phi$  for 
 every $\mathfrak{x}\in\R^4$, $\mathfrak{e}^{\cdot2} =1$, $e_0>0$, whence the claim.\qed
 
 \begin{Lem}\label{CCMTA}
 Let  $ \mathfrak{J}$  be conserved and covariant. Let  $\phi\in\mathcal{D}$. Suppose \emph{(\ref{CMTA})(a),(b)} for $\mathfrak{J}(\phi,\cdot)$ and suppose $\int J_0(\phi;0,x)\d^3x=||\phi||^2$. Then 
 $$ ||\phi||^2= \int \big( J_0(\phi;\tau(x),x) -  J(\phi;\tau(x),x)\nabla\tau(x)\big)  \d^3x  $$
 holds for every maximal achronal set with  corresponding $1$-Lipschitz function $\tau$. 
\end{Lem}

{\itshape Proof.} 
By covariance of $\mathfrak{J}$ regarding time translations it follows that $\mathfrak{J}(\phi;\tau(x),x)=\mathfrak{J}(W(-\tau(0))\phi;\tau(x)-\tau(0),x) =\mathfrak{J}(\phi';\tau'(x),x)$ for $\phi':=W(-\tau(0))\phi$, $\tau':=\tau-\tau(0)$. Hence the right side of (\ref{ETAU}) reads $\int J_0(\phi';0,x)\d^3x =||\phi'||^2=||\phi||^2$.\qed

The main result  follows.

\begin{The}\label{ALCCC} Let  the real bounded $C^1$ current $\mathfrak{J}$ be conserved covariant. Let $\phi\in\mathcal{D}$.  Let  $J_0$ satisfy 
\begin{itemize}
\item[\emph{(a)}] 
$J_0(\phi;0,x)\ge 0$ for $x\in\R^3$,
 $\int J_0(\phi;0,x)\d^3 x=||\phi||^2$, $J_0(\cdot\, ;0,x)$ is quadratic  for every $x\in\R^3$
\item[\emph{(b)}] $J_0(\phi,\mathfrak{x})\le C(1+|x|)^{-N}$ for $ |x|\ge|x_0|$ with some constants $N>3$ and   $C <\infty$ depending on $\phi$
\end{itemize}
Then there is a unique AL $T$  satisfying for every achronal Borel set $\Delta$
\begin{equation}
\langle \phi, T(\Delta)\phi\rangle= \int_{\varpi(\Delta)} \big( J_0(\phi;\tau(x),x) -  J(\phi;\tau(x),x)\nabla\tau(x)\big)  \d^3x \tag{1}
\end{equation}
where $\tau:\varpi(\Delta)\to \R$ with $\operatorname{graph}\tau=\Delta$. $T$ is covariant.
\end{The}\\
{\itshape Proof.} Uniqueness is obvious as $\mathcal{D}$ is dense. Let $\phi\in\mathcal{D}$. 
\\
\hspace*{6mm} 
By (a) and (\ref{FDP}), $J_0(\phi,\mathfrak{x})\ge |J(\phi,\mathfrak{x})|$.  
\\
\hspace*{6mm} 
Let $\Delta_0\in\mathcal{B}^{ach}$. There is   a maximal achronal set $\Lambda \supset \Delta_0$ being the graph of the corresponding  $1$-Lipschitz function  $\tau: \R^3\to\R$. For Borel $\Delta\subset\Lambda$ put
\begin{equation}
 \pi_{\phi,\Lambda}(\Delta):= \int_{\varpi(\Delta)} \big( J_0(\phi;\tau(x),x) -  J(\phi;\tau(x),x)\nabla\tau(x)\big)  \d^3x \tag{2}
\end{equation}
As the integrand  is nonnegative, $ \pi_{\phi,\Lambda}$ is a $\sigma$-additive measure. By (a) and (\ref{CCMTA}), $ \pi_{\phi,\Lambda}(\Lambda)=||\phi||^2$. Due to  (\ref{QDIAS}),\,(\ref{SPWQ}),
  $ \phi\mapsto \pi_{\phi,\Lambda}(\Delta)$ is the quadratic form of a bounded Hermitian sesquilinear form  on $\mathcal{D}$. Therefore by \cite[Lemma 48]{DM24} there is a bounded operator $T_{\Lambda}(\Delta)$, $0\le T_{\Lambda}(\Delta)\le I$ with $ \pi_{\phi,\Lambda}(\Delta)=\langle \phi, T_\Lambda(\Delta) \phi\rangle$.
\\
\hspace*{6mm}
Let $\phi\in\mathcal{H}$. {Approximating $\phi$ by $\phi_n\in\mathcal{D}$ finite additivity of  $\Delta\mapsto\langle \phi, T_\Lambda(\Delta) \phi\rangle$ follows by continuity.}  Actually it is $\sigma$-additive. Indeed, let $\Delta_n\downarrow_n\emptyset$ for Borel $\Delta_n\subset\Lambda$. Let $\epsilon>0$. Let $\phi'\in\mathcal{D}$ with $||\phi-\phi'||\le\epsilon$. An obvious application of the triangle inequality yields $$||\langle \phi, T_\Lambda(\Delta_n)\phi\rangle - \langle \phi', T_\Lambda(\Delta_n)\phi'\rangle|| \le ||\phi-\phi'|| \,||\phi||+||\phi'||\,        ||\phi-\phi'||\le 2\epsilon ||\phi||+\epsilon^2\le C\epsilon$$
with $C<\infty$ independent of $\Delta_n$. As  $\langle \phi', T_\Lambda(\Delta_n)\phi'\rangle \to 0$, one infers $\langle \phi, T_\Lambda(\Delta_n)\phi\rangle \to 0$, whence the claim. 
\\
\hspace*{6mm}
So $T_\Lambda$ is weakly $\sigma$-additive, which by  \cite[Theorem 4.28]{W76} implies the strong $\sigma$-additivity.
\\
\hspace*{6mm}
Note that the definition of $T_\Lambda(\Delta_0)$ via (2) does not depend on $\Lambda$. One may omit the index. Thus $T$ is an AL. It remains to show its covariance, which follows  immediately by the subsequent proposition (\ref{CPOL})(b).\qed

\begin{Rem} The assumption (\ref{ALCCC})\,(a) can be replaced by 
\begin{itemize}
\item[(a')]
$\int J_0(\phi;0,x) \d^3 x=||\phi||^2$ and $\langle \phi,T(\Delta)\phi\rangle=\int_\Delta J_0(\phi;0,x) \d^3 x$ with nonnegative operators $T(\Delta)$ for bounded Borel $\Delta\subset \R^3$ 
\end{itemize}
since, arguing as in (\ref{SPWQ}), (a') implies (a).
\end{Rem}

Regarding the notations see $(2)$ in the proof of (\ref{ALCCC}).

\begin{Pro}\label{CPOL} Let $g=(\mathfrak{a},A)\in \tilde{\mathcal{P}}$. Then 
\\
\hspace*{6mm}
\emph{(a)}  {the achronal set $g\cdot\Delta$ is given by}  $$g\cdot\Delta=\{(\tau_g(y),y):y\in \varpi(g\cdot\ \Delta)\}$$ for $\tau_g(y):=(g\cdot (\tau(x),x))_0$ with $x:=S^{-1}(y)$, where $$S:\R^3\to \R^3,\quad S(x):=\varpi \big(g\cdot (\tau(x),x) \big)$$ is a bijection.
\\
\hspace*{6mm}
\emph{(b)} $\pi_{W(g)^{-1}\phi,\Lambda}(\Delta)=\pi_{\phi,g\cdot\Lambda}(g\cdot\Delta)$.
\end{Pro}\\
{\it Proof.} (a) Obviously $S$ is surjective. Let $S(x)=S(x')$. Hence $$\varpi \big(g\cdot (\tau(x)-\tau(x'),x-x')\big)=0$$ with  
$\big(g\cdot (\tau(x)-\tau(x'),x-x')\big)^{\cdot 2}=\big(\tau(x)-\tau(x'),x-x'\big)^{\cdot 2}\le 0$. Therefore also $(g\cdot (\tau(x)-\tau(x'),x-x'))_0=0$, whence  $(g\cdot (\tau(x)-\tau(x'),x-x'))=0$. This means  $(\tau(x)-\tau(x'),x-x')=0$. So $x=x'$. In conclusion $S$ is bijective.
\\
\hspace*{6mm}
Note $S(\varpi(\Delta))=\varpi(g\cdot\Delta)$. Therefore $\{(\tau_g(y),y):y\in \varpi(g\cdot\ \Delta)\}=\{\big((g\cdot (\tau(x),x))_0, S(x)\big):x\in\varpi(\Delta)\} =\{g\cdot (\tau(x),x):x\in\varpi(\Delta)\}=g\cdot\Delta$.
\\
\hspace*{6mm}
(b) By (\ref{CCC})\,(ii),  $$\pi_{ W(g)^{-1}\phi,\Lambda}(\Delta)=
\int_{\varpi(\Delta) }\mathfrak{J}\big(\phi,g\cdot (\tau(x),x)\big)
 \cdot \big(A\cdot (1,\nabla\tau(x))\big)\d^3x$$  equals $ \int_{\varpi(\Delta) }\mathfrak{J}\big(\phi, \tau_g(S(x)),S(x)\big)
 \cdot \big(A\cdot (1,\nabla\tau(x))\big)\d^3x$ which using the image of the    Lebesgue measure  $\lambda$ becomes
 $$  \int_{S(\varpi(\Delta)) }\mathfrak{J}\big(\phi, \tau_g(y),y\big)
 \cdot \big(A\cdot (1,\nabla\tau(S^{-1}(y))\big)\d S(\lambda)(y)$$ Now recall  $S(\varpi(\Delta))=\varpi(g\cdot\Delta)$ and note $\d S(\lambda)/\d \lambda =|\det D\,S^{-1}|=|\det D\,S(S^{-1}(\cdot))|^{-1}$.
\\
\hspace*{6mm} 
It remains to verify    
\begin{equation}
(1,\nabla\tau_g(y))=   |\det D\,S(S^{-1}(y))|^{-1}\,   A\cdot (1,\nabla\tau(S^{-1}(y)))\tag{*}
\end{equation} 
which is easy in the case $A\in SU(2)$. So it suffices to check the case $g=\e^{\rho\sigma_3/2}$, $\rho\in\R$.\footnote{ Explicitly  $\e^{\,\rho\,\sigma_3/2}=\operatorname{diag}(\e^{\rho/2},\e^{-\rho/2})$ acts on $\R^4$ by $\left(\begin{array}{cccc}\cosh(\rho) & 0&0&\sinh(\rho)\\ 0&1&0&0\\0&0&1&0\\ \sinh(\rho)&0&0&\cosh(\rho)\end{array}\right)$ for $\rho \in \R$} Put $c:=\cosh \rho$, $s:=\sinh \rho$, $z:=\nabla \tau(x)$, $x=S^{-1}(y)$. The rows of  $ \big(D\,S(x)\big)^{-1}$ are $(1,0,0)$, $(0,1,0) $, $\frac{1}{c+sz_3}(-sz_1,-sz_2, 1)$. So the right side of (*) equals $$\frac{1}{c+sz_3}(c+sz_3,z_1,z_2,cz_3+s)\:.$$ On the left hand side $\nabla\tau_g(y))=\big(cz_1-\frac{(cz_3+s)sz_1}{c+sz_3},\dots,\frac{cz_3+s}{c+sz_3}\big)$. Hence (*) holds thus accomplishing the proof.\qed

\section{Covariant achronal localizations of the massive scalar boson}\label{sec6}

{The theory developed so far, in particular the construction of a covariant AL (\ref{ALCCC}),\,(\ref{CPOL}),  is rather general without requiring additional effort.
It refers to any relativistic quantum mechanical system, which is completely determined by a unitary representation $W$ of the Poincar\'e group. Here we turn to the elementary particle being the \textbf{massive scalar boson}, which is the  one-particle system
{of a free Klein-Gordon quantum field}
 uniquely determined by $W=[m,0,+]$, i.e., the mass $m>0$, the spin $0$ and the sign of energy $+$. In the momentum representation with $L^2(\R^3)$ being  the  space  of states $W$ reads\footnote{Often one uses the antiunitarily equivalent $ \e^{-\i \mathfrak{a}\cdot \,\mathfrak{p}}$.}
\begin{equation}\label{RMSB}
\big(W(\mathfrak{a},A)\phi\big)(p)=\sqrt{\epsilon(q)/\epsilon(p)}\, \e^{\i \mathfrak{a}\cdot \,\mathfrak{p}} \,\phi(q)
\end{equation}
with $\epsilon(p):=\sqrt{m^2+p^2}$,  $\mathfrak{p}:=(\epsilon(p),p)$, $\mathfrak{q}= (q_0,q):=A^{-1}\cdot \mathfrak{p}$.}
 \\
\hspace*{6mm}
One recalls that the localizability of the massive scalar boson in Euclidean space  is described  by  a Euclidean covariant  normalized POVM  $T$ on the Borel sets of $\R^3$, called a POL  (Positive Operator Localization) in \cite{C23}. 
 \\
\hspace*{6mm}
In the momentum representation by  \cite[(6.1), (11)\,Theorem]{C23} one has
\begin{itemize} 
\item $\langle \phi,T(\Delta)\phi\rangle=\int_\Delta J_0(\phi,x) \d^3 x$
\end{itemize} 
  i.e.,    (\ref{BRLC}) holds, where the density of the probability of localization $J_0$ is given by
\begin{itemize}
\item          $J_0(\phi,x) =(2\pi)^{-3}\int\int \, \textsc{k}(k,p) 
\e^{\i(p-k)x}\overline{\phi(k)}\phi(p)   \,\d^3k\,  \d^3p$
\end{itemize}
for  $\phi\in C_c$, i.e., continuous with compact support. 
Here  $\textsc{k}$ is any measurable normalized (i.e.,    $\textsc{k}(p,p)=1)$          rotational invariant positive definite  separable kernel  $\textsc{k}$ on $\R^3\setminus\{0\}$.
 \\
\hspace*{6mm}
The aim is to extend $T$ to an AL which is Poincar\'e covariant under the representation $W$ (\ref{RMSB}) of the massive scalar boson
following the considerations in sec.\,\ref{SALCCC}.  Petzold and collaborators \cite{GGP67} show that   $J_0$ is the zero component of a covariant  conserved four-vector current $\mathfrak{J}:=(J_0,J)$ if and only if

\begin{equation}\label{CCMR}
\mathfrak{J}(\phi,\mathfrak{x})=(2\pi)^{-3}\int\int \, \mathfrak{K}(k,p)
\e^{\i\big((\epsilon(k)-\epsilon(p))x_0-(k-p)x\big)}     \overline{\phi(k)}\phi(p)   \, \d^3k \d^3p
\end{equation}
with $\phi\in C_c$. Here 
\begin{equation}\label{FVCCD}
 \mathfrak{K}(k,p):=\frac{(\epsilon(k)+\epsilon(p), k+p)}{2\sqrt{\epsilon(k)}\sqrt{\epsilon(p)}} g\big(  \epsilon(k)\epsilon(p)-kp   \big)
 \end{equation}
  where
 $g: [m^2,\infty[ \to \R$ is  continuous with $g(m^2)=1$ such that  the zeroth component $K_0$ of $\mathfrak{K}$ is a positive definite kernel on $\R^3$ (see also \cite[(55)\,Corollary]{C23}). 

\begin{Def}\label{CK}
 $\mathfrak{K}$ in (\ref{FVCCD}) is called a \textbf{causal kernel} if its zeroth component is positive definite on $\R^3$.
\end{Def} \\
For a thorough analysis of the solutions $g$ see \cite{C23}. We mention  $|g(t)| < g_{3/2}(t)$ if $t\ne m^2$, $g\ne g_{3/2}$, where $g_r(t):=(2m^2)^r(m^2+t)^{-r}$  for $r\ge 3/2$ denotes the \textbf{basic series} of solutions revealed by \cite{GGP67} and \cite{HW71}.

\hspace*{6mm}
 Henceforth we deal with the conserved covariant  currents $\mathfrak{J}$  with causal kernel (\ref{CCMR}). For $\mathcal{D}:=C^\infty_c(\R^3)$ the assumptions on  $\mathfrak{J}$ in (\ref{CCC}) are satisfied.
$\mathfrak{J}(\phi, \cdot)$ is even smooth.  Moreover $J_0$ satisfies (\ref{BRLC}) and hence (\ref{ALCCC})(a) by (\ref{SPWQ}). Regarding the assumption (\ref{ALCCC})(b) one has

\begin{Lem}\label{C4G}   Let $\phi\in C^\infty_c(\R^3)$. For $g$ in \emph{(\ref{FVCCD})}  assume $g\in C^4([m^2,\infty[)$.
Then  \emph{(\ref{ALCCC})(b)}  holds.
\end{Lem}
\\
{\itshape Proof}. Let $x_0\ne 0$. Put $$F:\R^3\times\R^3\to \R, \quad F(k,p):=\frac{\varepsilon(k)x_0-kx}{|x|+|x_0|} - \frac{\varepsilon(p)x_0-px}{|x|+|x_0|}$$
 $F$ is $C^\infty$. Put $\varphi(k,p) := (2\pi)^{-3}K_0(k,p)\overline{\phi(k)}\phi(p)$. $\varphi$ is $C_c^4$. One has
\begin{equation}
J_0(\phi,\mathfrak{x})=\int\int \e^{\i (|x|+|x_0|)F(k,p)} \varphi(k,p)\d^3k\d^3p\tag{*}
\end{equation}
We proceed according the proof of \cite[Theorem 1.8]{T92}, which concerns  the non-stationary phase method.  A positive lower bound (**) of  $|\nabla F|$ is crucial.
\\
\hspace*{6mm}
Let $K:=\operatorname{supp}\phi$. Put $\beta:=\max\{\frac{|p|}{\varepsilon(p)}:p\in K   \}$. Clearly $0\le \beta <1$.
 Then $$\nabla_k F(k,p)=  (|x|+|x_0|)^{-1}\big(\frac{x_0}{\epsilon(k)}k-x\big)$$ {whence} $|\nabla_k F(k,p)|\ge  (|x|+|x_0|)^{-1}\big(|x|-|x_0|\frac{|k|}{\epsilon(k)}\big)\ge \frac{|x|-\beta |x_0|}{|x|+|x_0|}$ for $k\in K$.  Now assume $|x|\ge |x_0|$. Then $|\nabla_k F(k,p)|\ge \frac{1-\beta}{2}$  and 
 similarly  $|\nabla_p F(k,p)| \ge   \frac{1-\beta}{2}$       for $p\in K$. It follows
 \begin{equation}
 |\nabla F(k,p)|\ge \frac{1-\beta}{\sqrt{2}}>0 \; \text{ for } |x|\ge|x_0|, \; (k,p)\in K\times K\tag{**}
\end{equation}
Note also that the derivatives satisfy 
$|D_k^\alpha F(k,p)|\le 1$, $|D_p^\alpha F(k,p)|\le 1$ for $|\alpha|=1$, $|D_k^\alpha F(k,p)|\le |D^\alpha \varepsilon(k)|$,  $|D_p^\alpha F(k,p)|\le |D^\alpha \varepsilon(p)|$ for $|\alpha|\ge 2$. Moreover, $\operatorname{supp}\varphi\subset K\times K$.
\\
\hspace*{6mm}
Repeated integration by parts as in the proof of  \cite[Theorem 1.8]{T92} yields  
$$J_0(\phi,\mathfrak{x})=  (|x|+|x_0|)^{-n}  \int_{K^2} \e^{\i (|x|+|x_0|)F(v)}  \Phi^{(n)}(v)\d^6 v  \;\text{ for } |x|\ge|x_0|, \, n\le 4$$
using the notation $v:=(k,p)\in\R^6$ and $K^2=K\times K\subset \R^6$. Here the function $\Phi^{(n)}$ is a sum of products of factors $|\nabla F|^{-2}$, $D^\alpha F$ for $1\le|\alpha|\le 4$,  $D^\alpha\varphi$ for $|\alpha|\le 4$. Hence $\Phi^{(n)}$ is bounded on $K^2$ independent of $|x|,|x_0|$. The result follows.\qed
\\

Note that $g$ from the basic series $g_r(t)=  (2m^2)^r(m^2+t^2)^{-r}$, $r\ge 3/2$ is even $C^\infty$, in particular the distinguished $g=g_{3/2}$. 
 \\
  
One summarizes (\ref{CCMR}), (\ref{C4G}), (\ref{ALCCC}).

\begin{The}\label{FRAL} 
Let $\mathfrak{J}$ be a covariant conserved  current\footnote{recall (\ref{CCC}).} with causal kernel  for the massive scalar boson.  
Assume  $g\in C^4([m^2,\infty[)$. Then there is an  AL $T$ satisfying for every achronal Borel set $\Delta$ and  $\phi\in C^\infty_c(\R^3)$ 
\begin{equation*}
\langle \phi, T(\Delta)\phi\rangle= \int_{\varpi(\Delta)} \big( J_0(\phi;\tau(x),x) -  J(\phi;\tau(x),x)\nabla\tau(x)\big)  \d^3x 
\end{equation*}
where $\tau:\varpi(\Delta)\to \R$ with $\operatorname{graph}\tau=\Delta$. One has the covariance $$W(g)T(\Delta)W(g)^{-1}=T(g\cdot \Delta)\:.$$
\end{The}

We turn to the family of  localizations of the massive scalar boson obtained out of its  stress energy tensor \cite{M23}. For a thorough treatment see \cite[sec.\,6]{DM24}. The family is indexed  by the normalized future-directed timelike fourvectors $\mathfrak{n}$, i.e. $\mathfrak{n}^{\cdot 2}=1$, $n_0>0$. The related currents read still in the momentum representation with $\phi\in C_c^\infty(\R^3)$
\begin{equation}
\mathfrak{J}_\mathfrak{n}(\phi,\mathfrak{x})=(2\pi)^{-3}\int\int \, \mathfrak{K}_\mathfrak{n}(k,p)
\e^{\i\big((\epsilon(k)-\epsilon(p))x_0-(k-p)x\big)}     \overline{\phi(k)}\phi(p)   \, \d^3k \d^3p
\end{equation}

\begin{equation}
 \mathfrak{K}_\mathfrak{n}(k,p):=\frac{ \mathfrak{k}\cdot\mathfrak{n}\; \mathfrak{p}     + \mathfrak{p}\cdot\mathfrak{n}\; \mathfrak{k}  + (m^2- \mathfrak{k}\cdot\mathfrak{p}  )  \;\mathfrak{n}   }{2 \sqrt{\mathfrak{p}\cdot \mathfrak{n}}\: \sqrt{\mathfrak{k}\cdot \mathfrak{n}} \sqrt{\epsilon (p)} \sqrt{\epsilon (k)}} 
\end{equation}
with $\mathfrak{p}:=(\epsilon(p),p)$,  $\mathfrak{k}:=(\epsilon(k),k)$.
\\
\hspace*{6mm}
One easily checks that $\mathfrak{J}_\mathfrak{n}(\phi,\cdot)$ is real  smooth bounded conserved and that $J_{\mathfrak{n},0}(\cdot\,;0,x)$ is quadratic for every $x\in\R^3$. Also one verifies the covariance 
\begin{equation}\label{CFC}
 \mathfrak{J}_\mathfrak{n}\big(W(g)\phi,\mathfrak{x}\big)  = A\cdot  \mathfrak{J}_{A^{-1}\cdot \mathfrak{n}}(\phi,g^{-1}\mathfrak{x})              
 \end{equation}
for all $\mathfrak{n}$, $g=(\mathfrak{a},A)$, $\mathfrak{x}$. In addition,  for every $\mathfrak{n}$, $\phi$,  one has $J_{\mathfrak{n},0}(\phi,;0,\cdot)\ge 0$ and $\int J_{\mathfrak{n},0}(\phi,;0,x)\d^3 x=||\phi||^2$ as shown in \cite[(64), Theorem 54]{DM24}.   Lemma (\ref{FDP}),\,(\ref{QDIAS}),  and (\ref{CCMTA}) hold for every $\mathfrak{J}_{\mathfrak{n}}$ by the same proofs due to (\ref{CFC}). Finally, (\ref{ALCCC})(b) holds for every $J_{\mathfrak{n},0}$ by a proof analogous to that of (\ref{C4G}). In summary, (\ref{ALCCC}) applies to  $\mathfrak{J}_{\mathfrak{n}}$. It follows

\begin{The}\label{CFCLL} 
For every $\mathfrak{n}$ with $\mathfrak{n}^{\cdot2}=1$, $n_0>0$, there is an AL $M^\mathfrak{n}$ satisfying for every achronal Borel set $\Delta$ and  $\phi\in C^\infty_c(\R^3)$ 
\begin{equation*}
\langle \phi, M^\mathfrak{n}(\Delta)\phi\rangle= \int_{\varpi(\Delta)} \big( J_{\mathfrak{n},0}(\phi;\tau(x),x) -  J_\mathfrak{n}(\phi;\tau(x),x)\nabla\tau(x)\big)  \d^3x 
\end{equation*}
where $\tau:\varpi(\Delta)\to \R$ with $\operatorname{graph}\tau=\Delta$. One has the covariance $$W(g)M^\mathfrak{n}(\Delta)W(g)^{-1}=M^{g\cdot \mathfrak{n}}(g\cdot \Delta)\:.$$
\end{The}

\section{Covariant representation of the causal logic for the massive scalar boson}

The \textbf{causal logic} $\mathcal{C}$ is the lattice of Borel subsets of $\R^4$ which is partially ordered by  set inclusion $\subset$  and which is generated and orthocomplemented by 
 \textbf{achronal separateness}, i.e., the relation 
 \begin{equation}\label{AOR}
\mathfrak{x}\perp\mathfrak{y}\quad  \Leftrightarrow\quad  \mathfrak{x}\ne\mathfrak{y} \textnormal{ and } (\mathfrak{x}-\mathfrak{y})^{\cdot 2}\le 0 \:.
\end{equation}
The elements of $\mathcal{C}$ are called {\bf causally complete} sets. {In fact,  they satisfy the requirement $\Delta = (\Delta^\perp)^\perp$ where $M^\perp :=\{\mathfrak{x} \in \R^4 \: :\: \mathfrak{x} \perp \mathfrak{y}\:, \forall \mathfrak{y}\in M  \}$ is the {\bf causal complement} of $M \subset \R^4$, and $(M^\perp)^\perp$ is the {\bf causal completion} of $M$.}
\\
\hspace*{6mm} 
 The causal logic is thoroughly studied by Cegla, Jadczyk,
  in \cite{CJ77} and has been studied further on.
$\mathcal{C}$ enjoys some appealing physical properties similar to the ones of the lattice of abstract elementary propositions of a quantum system \cite{CJ77} and see also the more recent works \cite{C02,Further}:
it is  possible to prove that the lattice $\mathcal{C}$ is $\sigma$-complete, irreducible, orthomodular, atomic, atomistic exactly as for a quantum lattice, but fails to satisfy the {\em covering law} and separability. 
\\
\hspace*{6mm} 
Since  the work of \cite{CJ77} there persists the outstanding question how to construct covariant representations of the causal logic.

\begin{Def}\label{POC}
Let $F(M)$ for  $M\in \mathcal{C}$  be a bounded nonnegative  operator on $\mathcal{H}$. Suppose $F(\emptyset)=0$, 
and $\sum_nF(M_n)=I$ for every sequence $(M_n)$ of mutually orthogonal sets in $\mathcal{C}$ 
such that 
$\bigvee_nM_n=\R^4$. Then the map $F$ is  called a \textbf{representation of the causal logic} (RCL).
\\
\hspace*{6mm}
Let $W$ be a unitary representation of $\tilde{\mathcal{P}}$. Then   the RCL $F$ is said to be    (Poincar\'e)  \textbf{covariant}   by means of $W$ if $F(g\cdot M)=W(g) F(M)W(g)^{-1}$ holds for $g\in \tilde{\mathcal{P}}$ and $M\in\mathcal{C}$. 
\end{Def}\\
The convergence of the above sum occurs  in the weak operator topology (equivalently in the strong operator topology).

 By the way a RCL  $F$ is a map from the lattice  $\mathcal{C}$
	to the  {\em generalized $\sigma$ effect algebra of effects} $E(\mathcal{H})$ on $\mathcal{H}$  \cite{DP00}.  This latter enjoys a  weakened form of $\sigma$-complete orthocoplemented lattice structure. $F$ is a homomorphism of this weakened type of structure.
	In particular, 
it is easy to prove that a RCL  is $\sigma$-additive, order-preserving and orthocomplement-preserving. 

Only recently covariant RCL have been constructed. They concern  quantum mechanical systems with definite spin $j\in\N_0/2$ and mass spectrum 
$\subset$ $]0,\infty[$ of positive Lebesgue measure \cite{C24}. These results have been obtained by group theoretical methods.
\\
\hspace*{6mm}
A different approach is to relate RCL to Poincar\'e covariant conserved (operator) density currents. See the works cited  in  \cite{BJ79} and \cite{CJ79}. A first concrete step in realizing a representation is done in \cite{CJ79} essentially showing (\ref{ETAU}) for  a smooth conserved current with compact support.
\\
\hspace*{6mm}
Recently an other  idea of localization of the massive scalar boson  in causally complete regions is pursued  \cite{O24}. It
uses the method of modular localization from AQFT. Given a Cauchy surface,  via the  modular localization map, to every state and to the causal completion of every of its Borel subsets  a probability of localization is attributed. This is  asymptotically additive when increasing the distance between the regions of localization.
\\
\hspace*{6mm}
There is the  closed relationship  (\ref{EALCLL}) between AL and RCL, which is easy to expound. The set of \textbf{determinacy} of $M\subset\R^4$  is defined as
$$M^{\sim}:=\{\mathfrak{x}:\forall\;\mathfrak{z} \textnormal{ with }\mathfrak{z}^{\cdot 2}>0\;\exists\, s\in\R \textnormal{ with } \mathfrak{x}+s\mathfrak{z}\in M\}\:.$$
It consists of  all points $\mathfrak{x}$ such that every timelike  line through $\mathfrak{x}$  meets $M$. There is the remarkable result that, {if   $\Delta\in\mathcal{B}^{ach}$, then  $\Delta^\sim = (\Delta^\perp)^\perp$,  whence $\Delta^\sim\in\mathcal{C}$} and conversely, if $M\in\mathcal{C}$ then $M=\Delta^\sim$ for  $\Delta\in\mathcal{B}^{ach}$  maximal  achronal in $M$. For details see \cite{CJ77}, \cite{C17}, \cite{C24}.
 This feature of spacetime is the reason for the following one-to-one correspondence.

\begin{Pro}\label{EALCLL} See \emph{\cite[(19),\,(20)]{C24}}.
\\
\hspace*{6mm}
\emph{(a)} Let $F$ be an RCL. Set $T(\Delta):=F(\Delta^\sim)$ for $\Delta\in\mathcal{B}^{ach}$. Then $T$ is an AL. If $F$ is covariant by means of $W$, then so is $T$.
\\
\hspace*{6mm}
\emph{(b)}  Let  $T$ be an AL. Then there is a unique  RCL $F$ with $F(\Delta^\sim)=T(\Delta)$ for $\Delta\in\mathcal{B}^{ach}$. If $T$ is covariant by means of $W$, then so is $F$.
\end{Pro}

It suffices to join up (\ref{FRAL}),\,(\ref{EALCLL}).

\begin{The} To every causal kernel $\mathfrak{K}$ with  $g\in C^4([m^2,\infty[)$ there is a unique covariant\footnote{recall (\ref{POC})} RCL $F$  for the massive scalar boson such that $$\langle\phi,F(\Delta^\sim)\phi\rangle = \langle \phi, T(\Delta)\phi\rangle$$ holds for $\phi\in C^\infty_c(\R^3)$ and every achronal Borel set $\Delta$. Here $\langle \phi, T(\Delta)\phi\rangle$  is given in \emph{(\ref{FRAL})} with $\mathfrak{J}$ from 
\emph{(\ref{CCMR})}.
\end{The}\\
Thus, apparently for the first time, a covariant RCL for a quantum mechanical system with definite mass is achieved.
\\
\hspace*{6mm}
Similarly one obtains from (\ref{CFCLL}) a covariant family of RCL  related to the stress energy tensor of the massive scalar boson.

\begin{The} For every $\mathfrak{n}$ with $\mathfrak{n}^{\cdot2}=1$, $n_0>0$,  there is a unique RCL $F^\mathfrak{n}$ for the massive scalar boson such that  $$\langle\phi,F^\mathfrak{n}(\Delta^\sim)\phi\rangle = \langle \phi, M^\mathfrak{n}(\Delta)\phi\rangle$$ holds for $\phi\in C^\infty_c(\R^3)$ and every achronal Borel set $\Delta$. Here $\langle \phi, M^\mathfrak{n}(\Delta)\phi\rangle$  is given in \emph{(\ref{CFCLL})}. One has the covariance $W(g)F^\mathfrak{n}(\Delta)W(g)^{-1}=F^{g\cdot \mathfrak{n}}(g\cdot \Delta)$.
\end{The} 
\section{Final comments and QFT outlook}
All achievements presented in this work concern the notion of localization for a {particle in Minkowski spacetime, the massive scalar boson}.
We constructed very general notions of achronal  localization in complete agreement with requirements of Poincaré covariance and causality. It was thanks to 
suitable probability conserved currents and an appropriate use of the divergence theorem.  
\\
\hspace*{6mm}
{Pursuing  these investigations,  the methods developed here  will enable us to construct  the repeatedly tried and long-awaited representation of the causal logic for the Dirac electron and positron. The concept of localization in achronal regions  turns out to be also valid for the massless Weyl fermions. It gives rise to a causal localization and the equivalent representation of the causal logic. The remarkable feature of this localization is the existence of  maximal achronal regions and particle states with zero   probability of localization  within these regions.}
\\
\hspace*{6mm}
Referring to the discussion in sec.\,\ref{SALCCC}, the first type of causal   obstructions against any localization notion, historically  arising from Hegerfeldt's analysis and considered here into a more modern perspective started in \cite{C17}, were found to be definitely  harmless. {A fully feasible  notion} of achronal, causal, covariant localization is proved to be possible, provided one adopts the modern point of view where, in the general case, observables are described by (families of) POVMs  instead of (families of) projection-valued measures, namely selfadjoint operators. No sharply localized probability distributions are permitted in this approach. (Though sharply distributions can be approximated with arbitrary precision,  see \cite{C17,M23} and especially Sec.17 of \cite{C23}.) 
 It is interesting noting that, classic notions as the {\em Newton-Wigner position operator}, apparently ruled out by the Hegerfeldt theorem,    still keep to play a role in our framework   as the first-moment of the introduced families of POVMs as established in \cite{DM24} in a wide generality. This fact gives rise to the standard notion of localization in the rest space of an observer under suitable non-relativistic conditions (large mass limit).

Two types of probability currents, associated to respective notions of localization,   have been considered in this work for real massive  bosons (see sec.\ref{sec6}). {One type},  extensively studied   in \cite{C23} within a general framework, was  based on the theory of {\em causal kernels}. Another type discussed  in \cite{M23,DM24}, was constructed  in terms of the stress-energy tensor.

 The new perspective developed in this paper on the one hand refers to a completely general notion of rest {space \cite{C24} where} localization takes place. This notion is even more general than the one studied in \cite{DM24}, and is  represented by (maximal) achronal sets, using the weakest possible regularity condition explicitly required by physics. On the other hand, this generality permitted us to  construct, apparently for the first time, an effect-space  covariant representation of the lattice of  causally complete Borel sets of the spacetime, {\em the logic of spacetime},  for a boson particle with definite and strictly-positive  value of the mass. 
{Such a representation}, for particles with  unsharp mass spectrum was already {constructed in \cite{C24}.} {Achronal localization  and representation of the causal logic  for a quantum mechanical system are related by the fact that the probability of  localization  in an achronal region $\Delta$, which is 3D, is the same as in  its $4D$  causal completion  $\Delta^\sim$.} Or also in every other achronal region $\Delta'$ whose causal completion is the same $\Delta^\sim$.

From the perspective of QFT, our approach concerns the one-particle structure of the  Fock representation built upon the Poincaré invariant Gaussian vacuum state in Minkowski spacetime, for a massive real Klein-Gordon field.  As a consequence, at this stage of the study,  it is not directly  possible to implement a fully-fledged analysis of local causality in the Haag-Kastler perspective. An evident issue concerns the fact that  the projector onto the {one-particle Hilbert space} does not belong to a local algebra of observables. To overcome this obstruction it is necessary to see all introduced notions as restrictions to the one-particle space of more general notions defined  in terms of local algebras of the Klein-Gordon field.  The  restrictions should be considered as imposed by the choice of the used one-particle states.

Referring to the two types of covariant AL analyzed in Sec.\ref{sec6}, the one constructed from the stress–energy tensor \cite{DM24} in fact appears to be the one-particle space restriction of a similar notion defined on the full Fock space. This is because the stress–energy tensor operator used there is simply the restriction to the one-particle space of its Fock-space analogue. The covariant AL arising from causal kernels \cite{C23} requires a deeper analysis. However, at least in the case of a charged boson, the covariant AL studied in \cite{C23} should in some way be related to the one-particle–space restriction of the current operator defined in the Fock space.

The QFT perspective is the appropriate setting in which to attempt to address a second type of obstruction (see Sec.\ref{SALCCC}) to the existence of any notion of spatial localization. This issue, in addition to Hegerfeldt’s theorems, concerns the already mentioned theorems by Malament \cite{Malament} and Halvorson–Clifton \cite{HC}. These no-go results show that pairs of localization operators $T(\Delta)$ and $T(\Delta')$ (either orthogonal projectors or effects) generally do {\em not} commute even when they are associated with {\em causally separated} regions $\Delta$ and $\Delta'$. This was established under physically plausible assumptions, such as the additivity of $\Delta \mapsto T(\Delta)$ for $\Delta$ belonging to a given flat Cauchy surface, and the existence of a lower bound for the energy observable (the generator of time translations). The commutativity requirement as a mathematical expression of causal independence is assumed in \cite{HC} and, more generally, in the local-algebras formulation of quantum field theory \cite{Haag}, as a foundational condition. As a consequence, localization operators $T(\Delta)$, regardless of their definition, do not seem to satisfy a basic causality requirement. All these issues will be addressed elsewhere.

\section*{Acknowledgements}
V.M. and  C. De R. are grateful to Silvano Delladio and Andrea Marchese for useful suggestions and discussions, and acknowledge that this work has been written within the activities of INdAM-GNFM.
  
\section*{Declaration statements}

{\bf Conflict of Interest}: The authors declare that they have no conflict of interest.\\
{\bf Ethical Statement}: This work does not involve human participants, animals, or sensitive data, and therefore no ethical approval was required.\\
{\bf Informed Consent}: Not applicable.\\
{\bf Data Availability}: No datasets were generated or analyzed in this study. All relevant information is contained within the article.\\
{\bf Funding}: This research received no external funding.

\end{document}